\documentclass[a4paper,11pt]{article}
\pdfoutput=1
\usepackage{jcappub} 
\usepackage{newtxtext,newtxmath}

\usepackage[T1]{fontenc}

\usepackage{ae,aecompl}
\usepackage{graphicx,subfigure}	
\usepackage{amsmath}	
\usepackage{amssymb}	
\usepackage{multicol}        
\usepackage{bm}		
\usepackage{pdflscape}	
\usepackage{lineno}
\usepackage{tikz}
\usepackage{pagerange}

\usepackage[toc,page]{appendix}

\usepackage{graphicx}	
\usepackage{amsmath}	
\usepackage{amssymb}	

\usepackage{xspace}
\usepackage{footnote}
\usepackage{amsfonts,amsmath}
\usepackage[]{xcolor}
\usepackage{makecell}
\usepackage{srcltx}

\newcommand{\healpix}{\texttt{HEALPix}}

\newcommand{\gadget}{\texttt{gadget}}

\usepackage{etoolbox}

\usepackage[T1]{fontenc}
\usepackage{ae,aecompl}

\definecolor{mypink3}{cmyk}{0, 0.7808, 0.4429, 0.1412}

\usepackage{newtxtext,newtxmath}

\title{\boldmath{\fontsize{22}{11}\selectfont DEMNUni: The imprint of massive neutrinos on the cross-correlation between cosmic voids and CMB lensing}}
\author[a]{Pauline Vielzeuf}
\emailAdd{vielzeuf@cppm.in2p3.fr}
\author[b]{Matteo Calabrese}
\author[c]{Carmelita Carbone}
\author[d,e]{Giulio Fabbian}
\author[f,g,h]{and Carlo Baccigalupi}

\affiliation[a]{Aix-Marseille Univ, CNRS/IN2P3, CPPM, Marseille, France}
\affiliation[b]{Astronomical Observatory of the Autonomous Region of the Aosta Valley (OAVdA), Loc. Lignan 39, I-11020, Nus (Aosta Valley), Italy}
\affiliation[c]{INAF -- Istituto di Astrofisica Spaziale e Fisica cosmica di Milano (IASF-MI), Via Alfonso Corti 12, I-20133 Milano, Italy}
\affiliation[d]{Center for Computational Astrophysics, Flatiron Institute, 162 5th Avenue, 10010, New York, NY, USA}
\affiliation[e]{School of Physics and Astronomy, Cardiff University, The Parade, Cardiff, CF24 3AA, UK}
\affiliation[f]{SISSA, Via Bonomea 265, 34136 Trieste,  Italy}
\affiliation[g]{IFPU -- Institute for fundamental physics of the Universe, Via Beirut 2, 34014 Trieste, Italy}
\affiliation[h]{INFN -- Sezione di Trieste, via Valerio 2, 34127 Trieste, Italy}
\date{Accepted XXX. Received YYY; in original form ZZZ}

\abstract{Cosmic voids are a powerful probe of cosmology and are one of the core observables of upcoming galaxy surveys. The cross-correlations between voids and other large-scale structure tracers such as galaxy clustering and galaxy lensing have been shown to be very sensitive probes of cosmology and among the most promising to probe the nature of gravity and the neutrino mass. However, recent measurements of the void imprint on the lensed Cosmic Microwave Background (CMB) have been shown to be in tension with expectations based on LCDM simulations, hinting to a possibility of non-standard cosmological signatures due to massive neutrinos. In this work we use the DEMNUni cosmological simulations with massive neutrino cosmologies to study the neutrino impact on voids selected in photometric surveys, e.g. via Luminous Red Galaxies, as well as on the void-CMB lensing cross-correlation. We show how the void properties observed in this way (size function, profiles) are affected by the presence of massive neutrinos compared to the neutrino massless case, and show how these can vary as a function of the selection method of the void sample. We comment on the possibility for massive neutrinos to be the source of the aforementioned tension. Finally, we identify the most promising setup to detect signatures of massive neutrinos in the voids-CMB lensing cross-correlation and define a new quantity useful to distinguish among different neutrino masses by comparing future observations against predictions from simulations including massive neutrinos.}

\begin{document}
\label{firstpage}

\maketitle




\section{Introduction}

The upcoming generation of galaxy surveys, such as Euclid\footnote{\url{https://www.euclid-ec.org/}}, DES\footnote{\url{https://www.darkenergysurvey.org/}}, DESI\footnote{\url{https://www.desi.lbl.gov/}} or LSST\footnote{\url{https://www.lsst.org/}}, will map the recent universe on large sky fractions with unprecedented precision, and will constrain the cosmological model using, among other probes, the clustering of billion galaxies and their gravitational lensing. The correlation of these probes with observations of the Cosmic Microwave Background (CMB) anisotropies (through the Interaged Sachs Wolfe or Sunyaev-Zeldovich effect) as well as their gravitational lensing (CMB lensing) will also provide complementary source of information and additional constraining power on cosmology and systematics of the large-scale structures (LSS) probes~\cite[see e.g][ for some recent overview]{ilic2022,wenzl2022,zhang2022}. \\*
The imprint of LSS in CMB-based observables, being it that of  matter overdensities such as galaxy clusters and cosmic filaments~\cite{he2018} or  underdensities such as voids and troughs~\cite{gruen2016,brouwer2018}, has been detected several times in the past and compared to predictions of numerical simulations to assess its intrepretation. 
Since observations of neutrino oscillations~\cite{Becker-Szendy1992,Fukuda1998,ahmed2004} suggest the presence of at least two species of neutrinos with a non-zero mass, future experiments in cosmology (on the CMB and galaxy surveys side) will try to constrain the properties of massive neutrinos from their imprint at cosmological scales. As such, the effect of massive neutrinos on the whole set of cosmological observables analyzed by those experiments has to be carefully studied if reliable information has to be extracted from those probes.

\begin{figure}
    \centering
    \includegraphics[width=0.6\columnwidth]{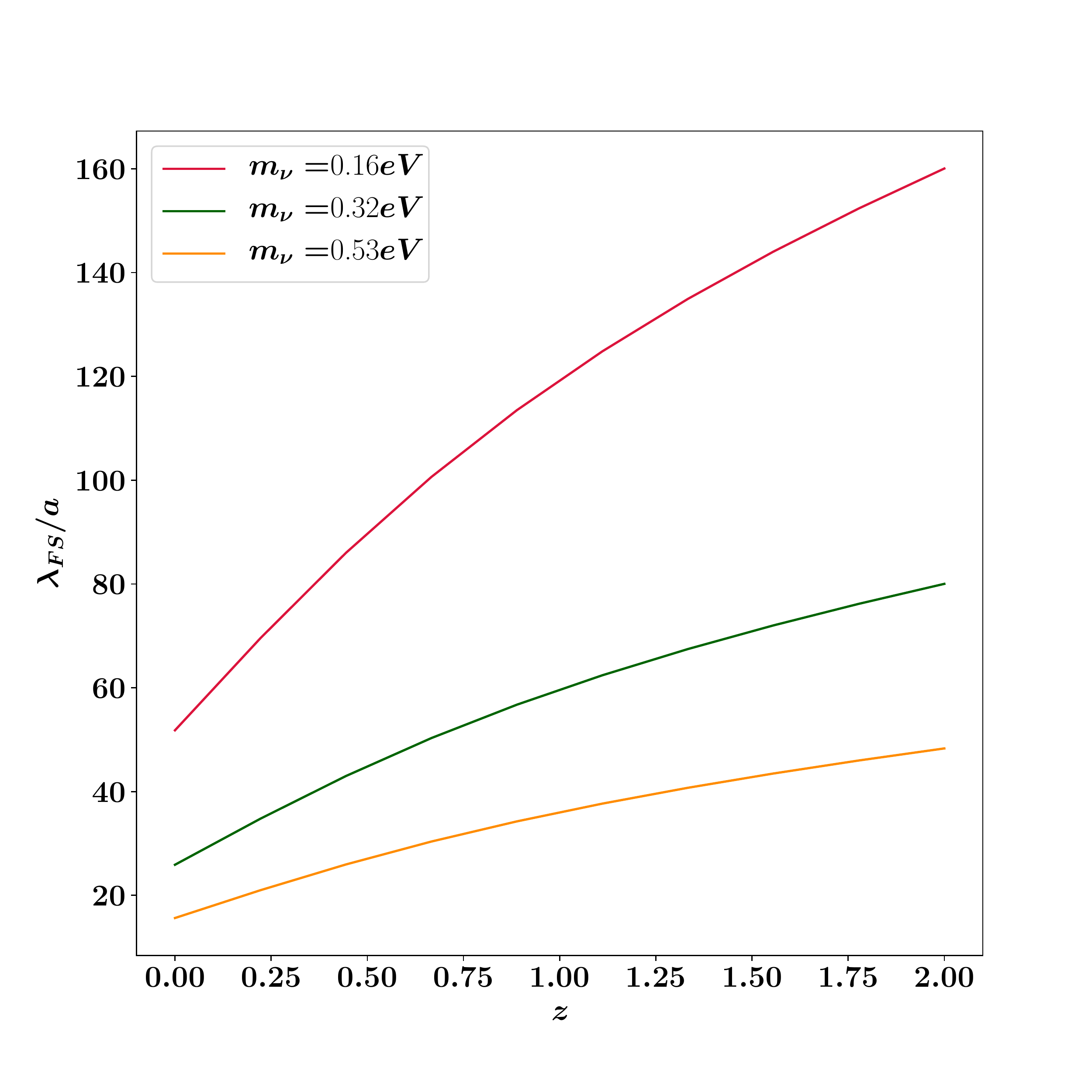}
    \caption{Comoving Free-Streaming length $\lambda_{\rm FS}$ of massive neutrinos - Eq.~\eqref{eq:FS} - as a function of redshift, for the different neutrino masses considered in this work.}
    \label{fig:FS}
\end{figure}

The presence of massive neutrinos in our universe has an impact in both background evolution and structure formation~\cite{Lesgourgues2006}. In particular, the evolution of the large structure in the cosmic web is directly sensitive to  the neutrino mass at scales of the order of the size of cosmic voids. Several  analyses have shown how cosmic voids could be exploited to set constraints on neutrino physics~\cite{Kreisch2019,bayer2021,contarini2021}. 
While at small scales, due to their non-zero velocity, massive neutrinos will travel across density fluctuations and effectively smooth them, at scales comparable to cosmic voids massive neutrinos will fall in the potential wells. We then expect cosmic voids to be particularly affected by the presence of massive neutrinos because the typical size of voids (10 to 100s of $h^{-1}$Mpc) approaches the free-streaming length ($\lambda_{\rm FS}$) of massive neutrinos, which can be expressed as function of redshift and neutrino species mass~\cite{Lesgourgues2012,Lesgourgues2013}:
\begin{equation}\label{eq:FS}
    \lambda_{\rm FS}(m_\nu,z)\sim 8.1 \frac{H_0(1+z)}{H(z)}\left(\frac{1eV}{m_\nu}\right)h^{-1}{\rm Mpc},
\end{equation}
with $H(z)$ and $H_0$ being the Hubble parameter and its value at $z=0$, respectively. As an example in Figure~\ref{fig:FS} we show the evolution of $\lambda_{\rm FS}$ as a function of redshift for the different neutrino masses that will be considered in this work.
Since the neutrino field is less clustered compared to the total matter field, the ratio between matter and neutrino density is higher at the maximum of the potential field, i.e. in cosmic voids. We thus expect to observe stronger effects of the presence of massive neutrinos in voids than in denser regions of the universe and~\cite{Zhang2019} has in fact shown that the halo mass function of dark matter haloes selected in cosmic voids is more affected by the presence of massive neutrino than by the full sample of the dark matter halo present in their volume. 
The overall presence of massive neutrinos therefore slows down the evolution of cosmic voids with respect to the massless neutrino case and induce modifications in the density profile of voids. We thus expect the signature of massive neutrinos to be also observable in the gravitational lensing effect~\cite{massara2015}. 


The deflection of photons (coming from CMB and background galaxies alike) induced by gravitational potentials along the line of sight tend to converge light rays onto denser regions and divert them away in the case of voids. Such {\it negative-lensing}  signal of cosmic voids has already been detected on the galaxy shape~\cite{clampitt2015,melchior2014,Carles_void,fang2019}, as well than in the CMB anisotropies and lensing 
\cite{cai,planck2016isw,vielzeuf2019,Raghunathan2019}. The most recent results from  DESI~\cite{hang2021} and DES~\cite{kovacs2022} have reported a lower signal in the observed lensing signal in the CMB at cosmic void positions w.r.t. the one estimated from $\Lambda$CDM simulations with massless neutrinos. Since the presence of massive neutrinos has been advocated as a way to reconcile such observational results with those observed in simulations, in this work we try to assess the validity of this hypothesis and study the gravitational lensing caused by cosmic voids modeled with N-body simulation with different neutrino mass on the CMB lensing field. 

This paper will be organized as follow, in Section~\ref{sec:DEMNUni} we will present the DEMNUni simulations halo samples and CMB-lensing (CMBL) maps used in this analysis, then in Sections~\ref{sec:void_finding} and \ref{sec:void_cat} the void finding methodology and the different void catalogues used in this analysis will be exposed. And finally Section~\ref{sec:CMBxvoids} will present the effect of massive neutrinos on the correlation of cosmic voids with CMB lensing signal.

\section{CMBL maps and halo lightcone}\label{sec:DEMNUni}

\subsection{The DEMNUni simulations}
The bulk of this work is based on numerical simulations from the ``Dark Energy and Massive Neutrino Universe'' 
(\href{https://www.researchgate.net/project/DEMN-Universe-DEMNUni}{DEMNUni})\cite{DEMNUni2}.
The DEMNUni simulations have been produced with the aim of investigating large-scale structures in the presence of massive neutrinos and dynamical dark energy, and they were conceived for the nonlinear analysis and modelling of different probes, including dark matter, halo, and galaxy clustering~\cite{DEMNUni1,Moresco2017,Zennaro2018,Ruggeri2018,Bel2019,Parimbelli2021,Parimbelli2022, Baratta_2022,Guidi_2022, SHAM-Carella_in_prep}, weak lensing, CMB lensing, SZ and ISW effects~\cite{Roncarelli2015,DEMNUni_simulations,fabbian2018, Beatriz_2023}, cosmic void statistics~\cite{Kreisch2019,Schuster2019,verza_2019,Verza_2022a,Verza_2022b}, and cross-correlations among these probes~\cite{Cuozzo2022_inprep}.
The DEMNUni simulations combine a good mass resolution with a large volume to include perturbations both at large and small scales. They are characterised by, a softening length $\varepsilon=20\, h^{-1}$ kpc, a comoving volume of $(2 \: h^{-1}\mathrm{Gpc})^3$ filled with $2048^3$ dark matter particles and, when present, $2048^3$ neutrino particles. The simulations are initialised at $z_{\rm in}=99$ with Zel'dovich initial conditions. The initial power spectrum is rescaled to the initial redshift via the rescaling method developed in~\cite{zennaro_2017}. Initial conditions are then generated with a modified version of the \texttt{N-GenIC} software, assuming Rayleigh random amplitudes and uniform random phases.
The DEMNUni simulations were run using the tree particle mesh-smoothed particle hydrodynamics (TreePM-SPH) code \gadget{} \citep{Springel_2005}, specifically modified as in \citep{Viel_2010} to account for the presence of massive neutrinos. This modified version of \gadget{} follows the evolution of cold dark matter (CDM) and neutrino particles, treating them as two separated collisionless components. For each simulation we have produced 63 output logarithmically equispaced in the scale factor $a=1/(1+z)$, 49 of which lay between $z=0$ and $z=10$.
The DEMNUni suite accounts for 15 different cosmological models with different neutrino masses and dynamical dark energy. However, in this work we focus on four separate numerical simulations: one in a standard neutrino massless $\Lambda$CDM model, and three in a modified $\Lambda$CDM cosmology characterised by the presence of massive neutrinos with total mass $m_{\nu}=$ $0.16$ eV, $0.32$ eV, $0.53$ eV. All these simulation share the same baseline, {\it Planck-2013} cosmology \citep{Planck2013_XVI}:
\begin{equation*}
\{ \Omega_{dm},  \Omega_{b},  \Omega_{\Lambda}, n_s, \sigma_8, H_0
  \}  = \{ 0.27, 0.05, 0.68, 0.96, 0.83, 67 \: \rm{ Km / s /
    Mpc} \}. 
\end{equation*}

In this work we use the friend-of-friend (FoF) halo catalogues, built from each of the 63 particle snapshots via the  FoF algorithm included in \gadget{}~\citep{springel01,dolag09}, setting to 32 the minimum number of CDM particles, thus fixing the halo minimum mass to $m_{\rm FoF}\simeq2.5\times 10^{12}h^{-1}M_\odot$.

\subsection{CMBL reconstruction}
\label{sec:lightcone}

The lensing observables maps are extracted with a post-processing procedure acting on the N-body particle snapshots to create a full {\it lightcone}. This procedure follows the approaches of \citep{fosalba08,dasbode2008}, and was developed to perform high-resolution CMB lensing simulations \citep{fabbian2018,Hilbert2020}, in order to implement a multiple-lens ray-tracing algorithm in spherical coordinates on the full sky. 

The current version of the code reconstructs a full-sky, backward lightcone around an observer  using the particle snapshots out to the comoving distance of the highest redshift available from the simulation, following \cite{calabrese2015}. To overcome the finite size of an N-body simulation box, the code replicates the box volume the number of times in space necessary to fill the entire volume between the observer and the source plane. Moreover, the code can randomize the particle positions, as described in \cite{carbone2008, carbone2009}, throughout flips and shifts, to minimize any numerical artifacts due to the repetition of the same structures along the line of sight.
The backward lightcone is then sliced into 62 full-sky spherical shells with the following scheme: the median comoving distance spanned by each shell coincides with the comoving distance at the redshift of each N-body snapshot. Any particle inside each of these shells is projected onto 2D spherical maps; the resulting surface mass density $\Sigma$ on each sphere is thus defined on a two-dimensional grid. For each pixel of the $i$-th sphere one has
\begin{equation}
\Sigma^{(i)}(\boldsymbol{\theta}) = \frac{n \; m_{X}}{A_{\rm pix}}\,,
\label{eq:surfmass}
\end{equation}
where $n$ is the number of particles in the pixel, $A_{\rm pix}$ is the pixel area in steradians and $m_X$ is the particle mass of type $X$ (dark matter, or neutrino) from the N-body simulation. 
For this work, the algorithm produces for each N-body simulation a full-sky convergence map on a \healpix{}\footnote{\url{http://healpix.sourceforge.net}} grid \citep{Hp} with $n_\text{side} = 4096$, which corresponds to a pixel resolution of $0.85$ arcmin. 
Finally, the lensing convergence of a source plane at redshift $z_S$ is computed in the Born approximation as the weighted sum of surfaces mass density:
\begin{equation}
\label{eq:kappa}
\begin{split}
 \kappa (\boldsymbol{\theta}, \chi_S) &= 
\frac{4 \pi G}{c^2}\frac{1}{f_S}\!
\sum_{i} 
(1 + z_D^{(i)}) \frac{f_{DS}^{(i)}}{f_D^{(i)}}
\left[\Sigma^{(i)}( \boldsymbol{\theta})\!-\!\bar{\Sigma}^{(i)}\right].
\end{split}
\end{equation}
Eq. \eqref{eq:kappa} follows the standard notation in the literature of weak lensing observables \citep[see, e.g.,][ for reviews]{2001PhR...340..291B,2015RPPh...78h6901K, 2018ARA&A..56..393M}, for the convergence field $\kappa$ at an angular position $\boldsymbol{\theta}$ of a source at comoving line-of-sight distance $\chi_S$ and redshift $z_S = z(\chi_S)$. Note that $f_{DS} = f_K(\chi_S - \chi_D)$, $f_D = f_K(\chi_D)$ and $f_S = f_K(\chi_S)$, where $f_K(\chi)$ denotes the comoving angular diameter distance for comoving  line-of-sight distance $\chi$, and $z_D = z(\chi_D)$ is the redshift corresponding to comoving line-of-sight distance $\chi_D$. Finally, $\Sigma^{(i)}$ represents the angular surface mass density, while $\bar{\Sigma}^{(i)}$ is the mean angular surface mass density of the $i$-th shell; $f_{DS}^{(i)}$ and $f_D^{(i)}$ are the corresponding distances at the redshift of the $i$-th shell. $\Sigma^{(i)}$ is extracted directly from the N-body simulation with the map-making procedure described before. The angular position of the centre of each \healpix{} pixel coincides with the direction of propagation of the rays in the Born approximation. Several lensing observables can be then constructed by changing the source distance (or redshift $z_S$), which will affect the geometrical weight in the sum of Eq. \eqref{eq:kappa}. 

\begin{figure}[htbp]

\includegraphics[width=1.\columnwidth]{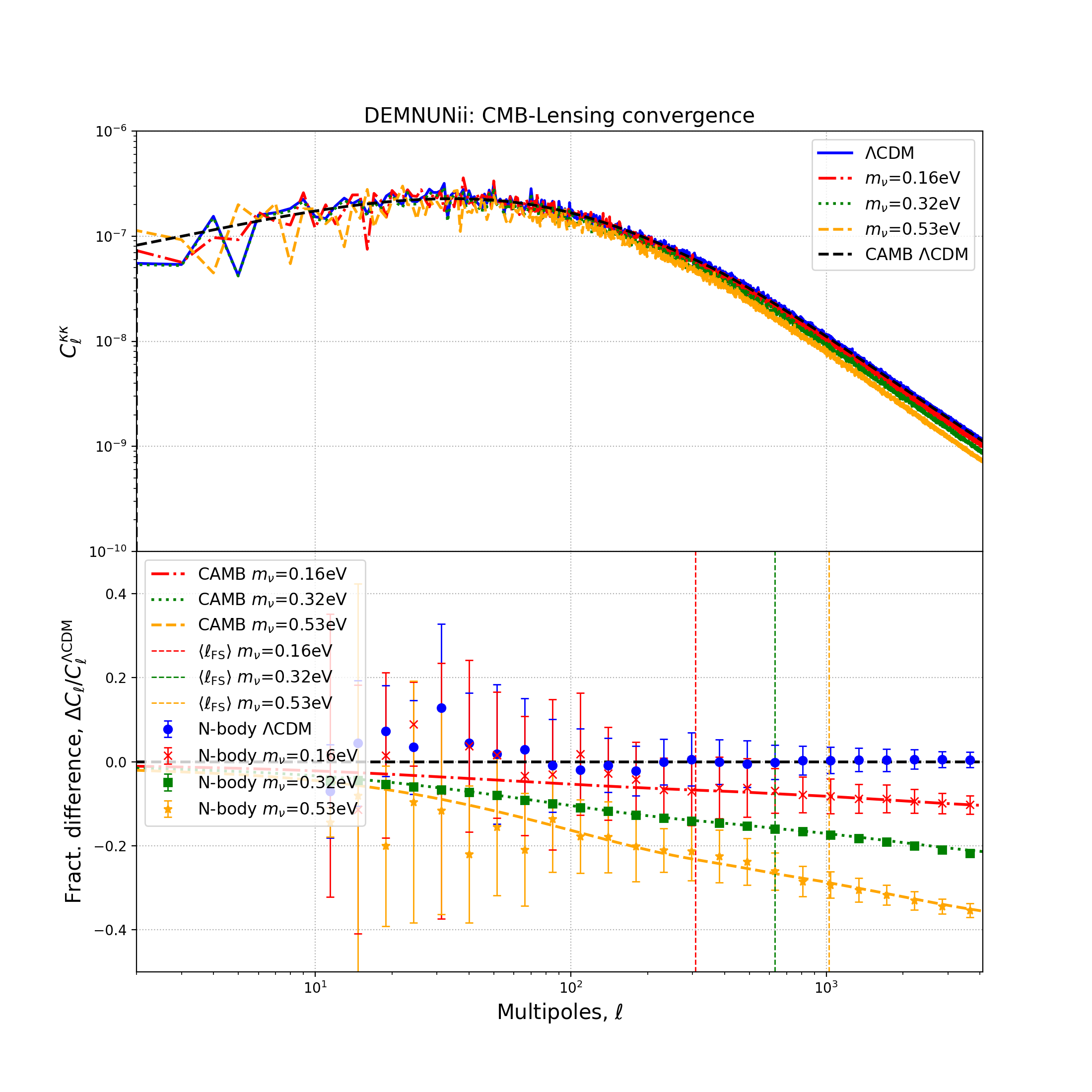}

\caption{{\it Top panel:} CMB convergence angular power spectrum,  for $\Lambda$CDM (blue line) and $\Lambda{\rm CDM} + m_\nu$ simulations with total neutrino masses $m_\nu=0.16$ eV (red, dot-dashed line), $m_\nu=0.32$ eV (green, dotted line) and $m_\nu=0.53$ eV (orange, dashed lines). Black, dashed line is the semi-analytical realization with \texttt{pyCAMB} for the DEMNUni $\Lambda$CDM cosmology. {\it Bottom panel:} fractional difference for the angular power spectra with respect to the $\Lambda$CDM case. Points with errorbars refer to measurements from N-body simulations via the lightcone convergence maps; signals have been binned in multipoles, error bars representing the variance in each bin. Lines are semi-analytical realizations with \texttt{pyCAMB} in the different cosmologies. Vertical lines are the (average) FS multipole - $\langle \ell_{\rm FS} \rangle$ - as computed by Eq.~\eqref{eq:elleFS}.}

\label{fig:cls_kcmb_dem}
\end{figure}

In this work, we have constructed convergence maps for CMB lensing ($\kappa$CMB), i.e. setting the source plane at the last scattering surface ($z_S$=1089). Four lightcones have been produced using particles snapshots in the four different cosmological scenarios: $\Lambda$CDM, and $\Lambda{\rm CDM} + m_\nu$ simulations with total neutrino masses $m_\nu=0.16$ eV, $m_\nu=0.32$ eV  and $m_\nu=0.53$ eV. In Figure~\ref{fig:cls_kcmb_dem} we show the angular power spectra from CMB convergence maps, both in terms of actual spectra (top panel) and ratio w.r.t. the $\Lambda$CDM case (bottom panel). Besides, signals are also compared with the semi-analytical realizations of \texttt{pyCAMB}\footnote{\url{https://camb.readthedocs.io/en/latest/}}, for all the different cosmologies considered. The bottom panel of this Figure highlights the effects of neutrino masses on the spectra, reducing the power especially at high multipoles where neutrino physics is more relevant. Points with errorbars are measurements on the reconstructed lightcone convergence ($\kappa$CMB) maps, while lines are semi-analytical realizations with \texttt{pyCAMB}. N-body simulations - and their related lightcone maps - are in line with theoretical expectations, especially for multipoles $\ell > 50$, whereas larger scales are affected by cosmic variance as seen in the top panel. Vertical lines in the Figure are computed directly from the free-streaming length of the different neutrinos masses considered - see Eq. \eqref{eq:FS} - translated into an (average) multipole $\langle \ell_{\rm FS} \rangle$, which represents the scale where neutrino effects could become more relevant. Multipoles are averaged on the CMB weak-lensing kernel (basically, the geometrical weight of Eq. \eqref{eq:kappa}), as:
\begin{equation}\label{eq:elleFS}
    \langle \ell_{\rm FS} \rangle = \frac{\int_{0}^{z_{\rm survey}}W^{\kappa \rm CMB}(z)\, 2\pi / \lambda_{\rm FS}(z) \chi(z) {\rm d}z}{\int_{0}^{z_{\rm survey}}W^{\kappa \rm CMB}(z) {\rm d}z};
\end{equation}
note how the free-streaming length is converted into multipole by the Limber approximation, $\ell_{\rm FS}  = 2 \pi / \lambda_{\rm FS} \cdot \chi$, and $z_{\rm survey}=2.0$ as the maximum redshift of the halo and void catalogues. The CMB lensing kernel is a function of the redshift of the lensed object, and can be expressed as follows:
\begin{equation}\label{eq:cmblensingkernel}
    W^{\kappa \rm CMB}=\frac{3 \Omega_m H_0^2}{2c}\frac{1+z}{H(z)}\chi(z)\frac{\chi(z_{\rm CMB})-\chi(z)}{\chi(z_{\rm CMB})}, 
\end{equation}
with the kernel peaking at $z \sim 1.5$. 

\subsection{The halo catalogue: construction and  measurements}\label{sec:hmf}
The halo sample is constructed using the same lightcone prescription for the lensing convergence maps, as described in the previous Section. The FoF sample extracted from the DEMNUni simulations is the basis for a full-sky, 3D-halo catalogue, where each object is placed around a central observer. In this case, each FoF halo behaves as a (CDM or neutrino) particle in the lensing lightcone, i.e. they are replicated a number of times in space necessary to fill the entire volume, which encompasses several spherical slices representing the Universe at different stages of its evolution. These slices follow the prescription and randomisation procedure described in Section~\ref{sec:lightcone} for CMB lensing. Therefore, we have produced for each considered cosmology a full-sky, 3D halo catalogue where each object is defined by its coordinates ($r, \theta, \phi$), where $r$ is the distance from the central observer at $z=0$ (thus $r$ it is also a measure of redshift), and $\theta$, $\phi$ are the standard coordinates on the celestial sphere. This procedure is then applied to all N-body simulations, building halo catalogues in four different cosmological scenarios.

\begin{figure}[htbp]
\begin{center}
\includegraphics[width=.8\columnwidth]{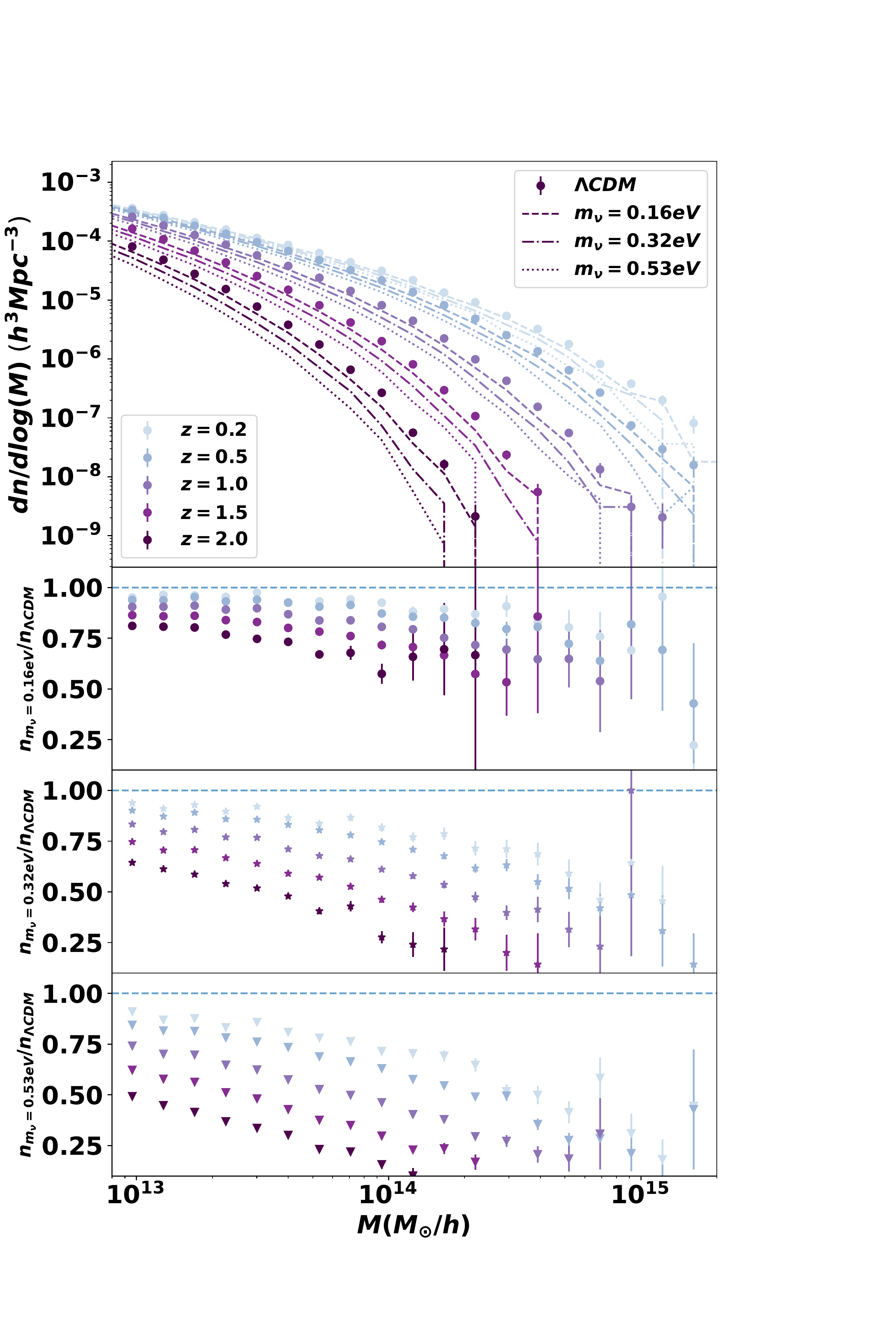}
\end{center}
\caption{{\it Top panel:} Halo mass function in redshift bins for both $\Lambda$CDM (error bars) and $\Lambda{\rm CDM} + m_\nu$ simulations with the three degenerate neutrino masses studied here, $m_\nu=0.16$ eV (dashed line), $m_\nu=0.32$ eV (dash-dotted line) and $m_\nu=0.53$ eV (dotted lines). {\it Bottom panels:} ratio of the halo mass function in massive neutrino cosmologies w.r.t. the $\Lambda$CDM case. Errorbars are derived assuming a Poisson distribution in each of the mass bins.} 

\label{fig:halo_mass_dist}
\end{figure}

Figure~\ref{fig:halo_mass_dist} shows the DM halo mass function measured on the DEMNUni halo catalogue for both massive and massless neutrinos cosmologies. The Figure highlights how the presence of massive neutrino tends to lower the halo abundance, and that this effect is more pronounced as the mass of the neutrinos increases. This is expected, since the free-streaming of massive neutrino induces counter effects to gravitational collapse and therefore a slow-down of the growth of structures. Moreover, in agreement with previous works (see for e.g. \cite{Brandbyge2010,marulli2011,Castorina}), this suppression is particularly evident for the heaviest haloes ($M_h>10^{14}h^{-1}\, M_\odot$). Bottom panels of Figure~\ref{fig:halo_mass_dist}, in fact, show the relative ratio of the halo mass function in the presence of massive neutrino w.r.t. the case where the neutrino are massless, per redshift bin and mass bin. In particular, we notice that this effect is stronger as the redshift increases: as we go to higher redshift the free-streaming scale of massive neutrinos increases, and consequently the presence of massive neutrinos will smooth the matter field at higher scales.

The effect of massive neutrinos in the halo mass function (and more generally in the matter clustering) will have an impact on the void population identified in the next step of the analysis. On the one hand, as shown in Figure~\ref{fig:halo_mass_dist}, the number of DM haloes formed in the presence of massive neutrinos decreases as the mass of the neutrinos increases, for the fixed minimum mass of the DEMNUni simulations; thus, the density of the considered tracers identifying voids will decrease as the neutrino mass increases. On the other hand, since the presence of massive neutrino slows down the clustering process, the trend is the opposite for haloes with smaller minimum mass, which can have densities higher in massive neutrino cosmologies than in the massless neutrino case \cite{liu2018}. Nonetheless from an observational point of view, halo number density of the DEMNUni simulations is consistent with galaxy densities from future survey observation as Euclid and LSST.

The decrease in the number of matter tracers per unit volume has a direct impact on the size distribution of voids: by using a sparser tracer sample, one should expect the resulting catalogue to contain fewer small structures, that will be eventually merged in larger ones \cite{sutter2014}\footnote{Note that while this trend can be seen for voids identified in the halo field, it can be different if CDM particles are used as void tracers (see \cite{massara2015} as an example).}. On the other hand, the fact that the clustering of the matter will be less effective due to massive neutrinos implies that the distribution of the underlying matter field will be sparser, and we therefore can expect to find shallower and larger underdensities w.r.t. the standard massless neutrino case.

\section{The void lightcone}

\subsection{The void finder}
\label{sec:void_finding}
During the past decades, various methodologies have been developed to identify underdensed regions of the matter field (see for example a comparison of void finders in~\cite{Colberg2008}). In the context of massive neutrino cosmologies, mainly 3-dimensional voids finders which implement Voronoi tessellation techniques have been  applied. In particular,~\cite{massara2015,Kreisch2019,Schuster2019} have quantified the effect of different massive neutrino cosmology on cosmic void properties and statistics using the DEMNUni simulations with the \texttt{VIDE}~\cite{sutter2015} 3-dimensional definition of voids. However, in~\cite{cautun2018} the authors have shown the potential of voids identified in 2-dimensional projected density fields to increase the detection level of void-lensing signals. 
In~\cite{Carles_void} - see also~\cite{vielzeuf2019,kovacs2019,kovacs2020,fang2019,kovacs2022} - the authors presented a void finder that identifies 2-dimensional underdensities in the matter field. Specifically, this void finder has been developed in the context of photometric observations in order to encompassed the photometric redshift errors, and it has been shown to optimize the strength of the lensing signal at void positions. Such results are in agreement with~\cite{cautun2018}, which argued that elongated structures on the line of sight (such are 2D voids by definition) will have a stronger lensing signal than spherical, 3 dimensional voids. Thus in this analysis, we have used this 2D void finder to identify cosmic voids in the halo catalogs from the DEMNUni simulation presented in Sect.~\ref{sec:DEMNUni}. The void finder can identify underdense regions in a tracer field following these steps:

\begin{enumerate}
    \item Divide the tracer sample in redshift slices, of a predefined comoving size $s_v$. In the context of this work, we have followed~\cite{Carles_void,vielzeuf2019} where the slice size is set to $s_v= 100 h^{-1}$Mpc, a value that has been shown to minimize error on redshift estimation in photometric survey;
    \item Convert each redshift slice in an \healpix{} map, and count the number of tracers located in each pixel of the map;
    \item Smooth the \healpix{} map with a gaussian kernel, the smoothing scale being left as free parameter of our methodology; 
    \item Identify the most underdensed pixel in the density map; 
    \item Compute the density in concentric shells around the void centre in the pixelized density maps until the shell density reaches the mean density of the slice itself; 
    \item Repeat the process for the next most underdensed pixels in the slice.
    
\end{enumerate}

The output catalogue will contain information on different voids parameters such as the void radius $R_{\rm v}$ - the radius at which the void density reaches the mean density of the slice - the mean void density $\Bar{\rho}$ - the averaged value of all pixel located inside the void radius - and the central density $\rho_{1/4}$, being the density of the central part of the identified void ($r<\frac{1}{4}R_{\rm v}$). These features in the output catalogue will depend on the different input parameters ({\it e.g.} matter tracer density, smoothing parameter, etc.).

\subsubsection{Void finder parameters}
\begin{figure}
\begin{center}
\includegraphics[width=.4\columnwidth]{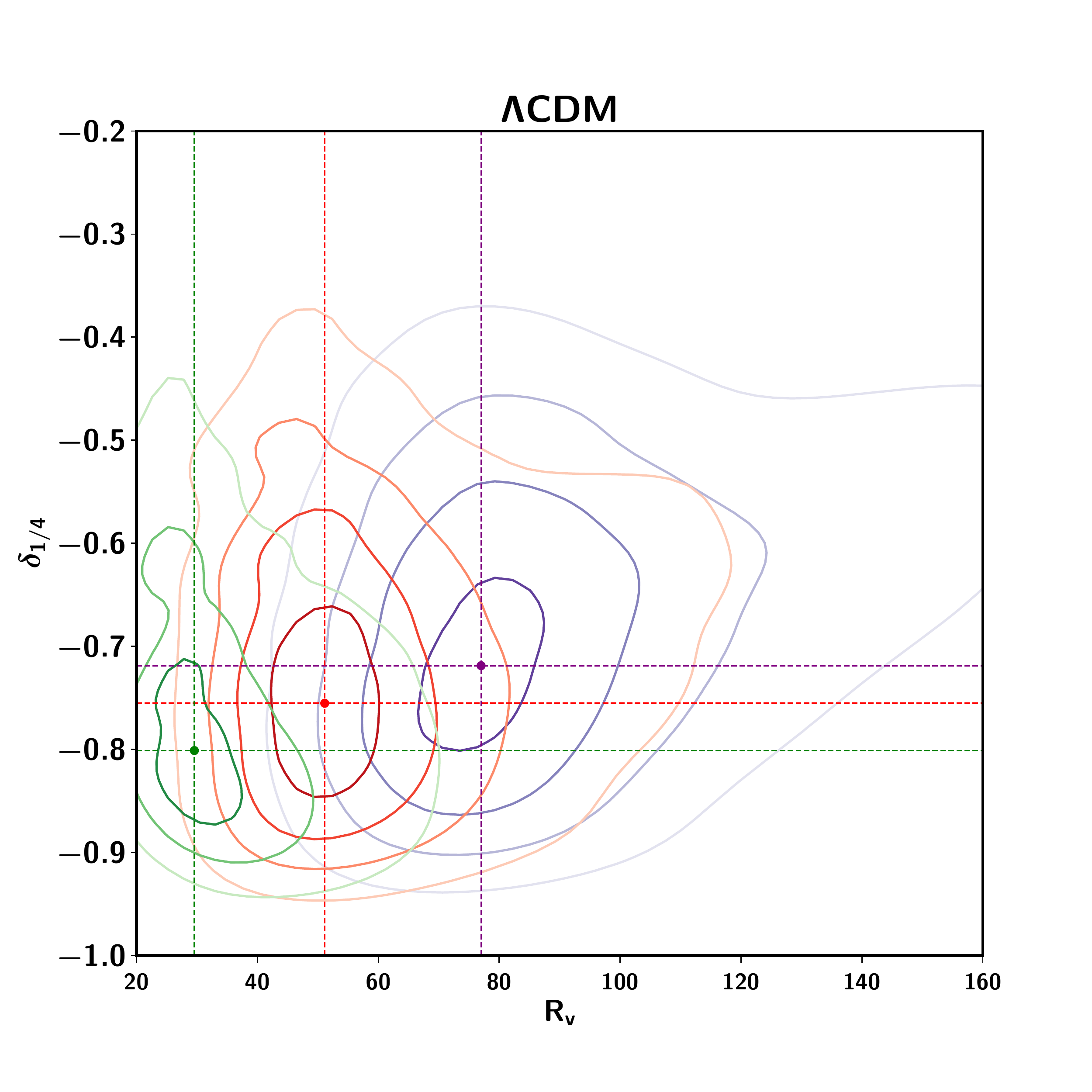}
\includegraphics[width=.4\columnwidth]{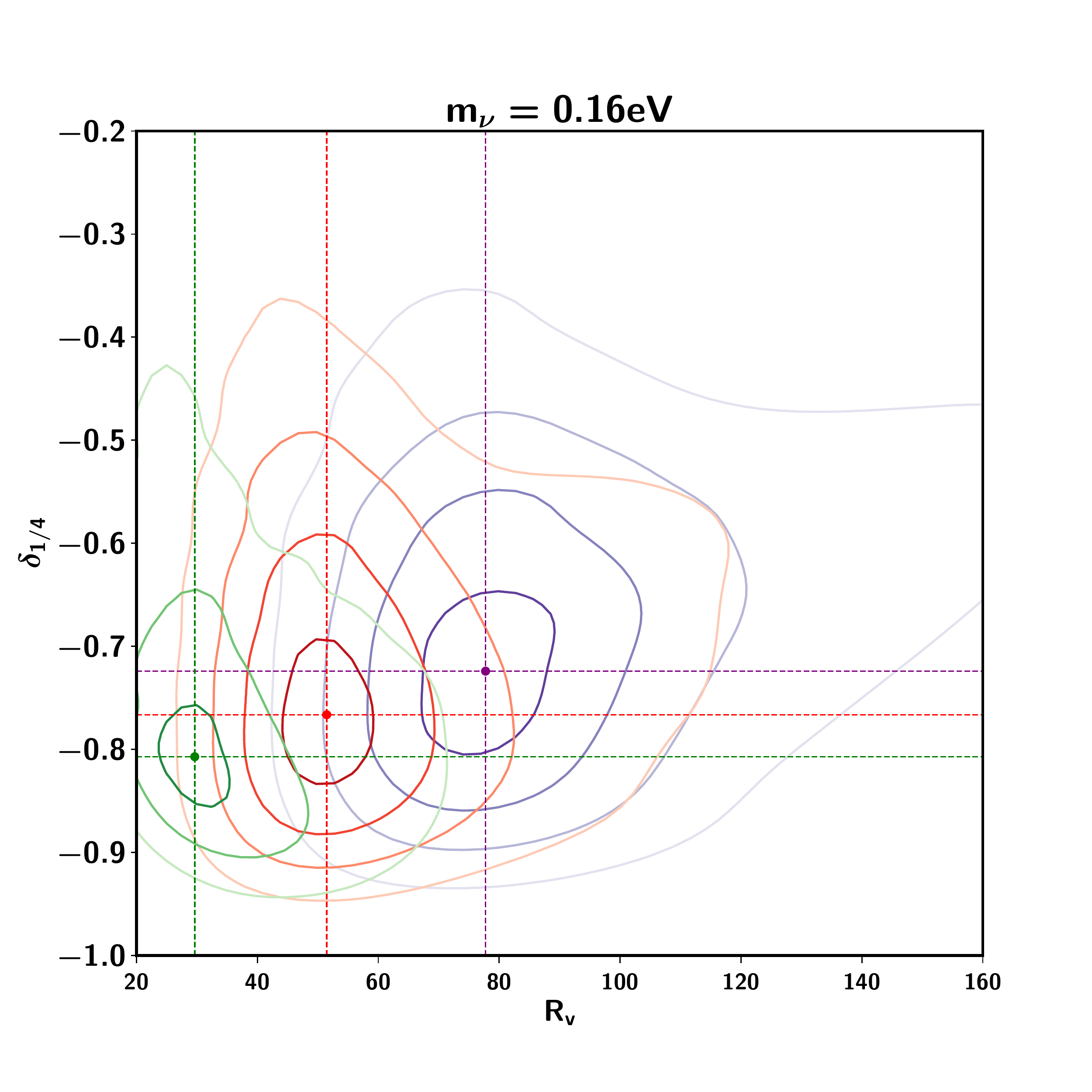}
\includegraphics[width=.4\columnwidth]{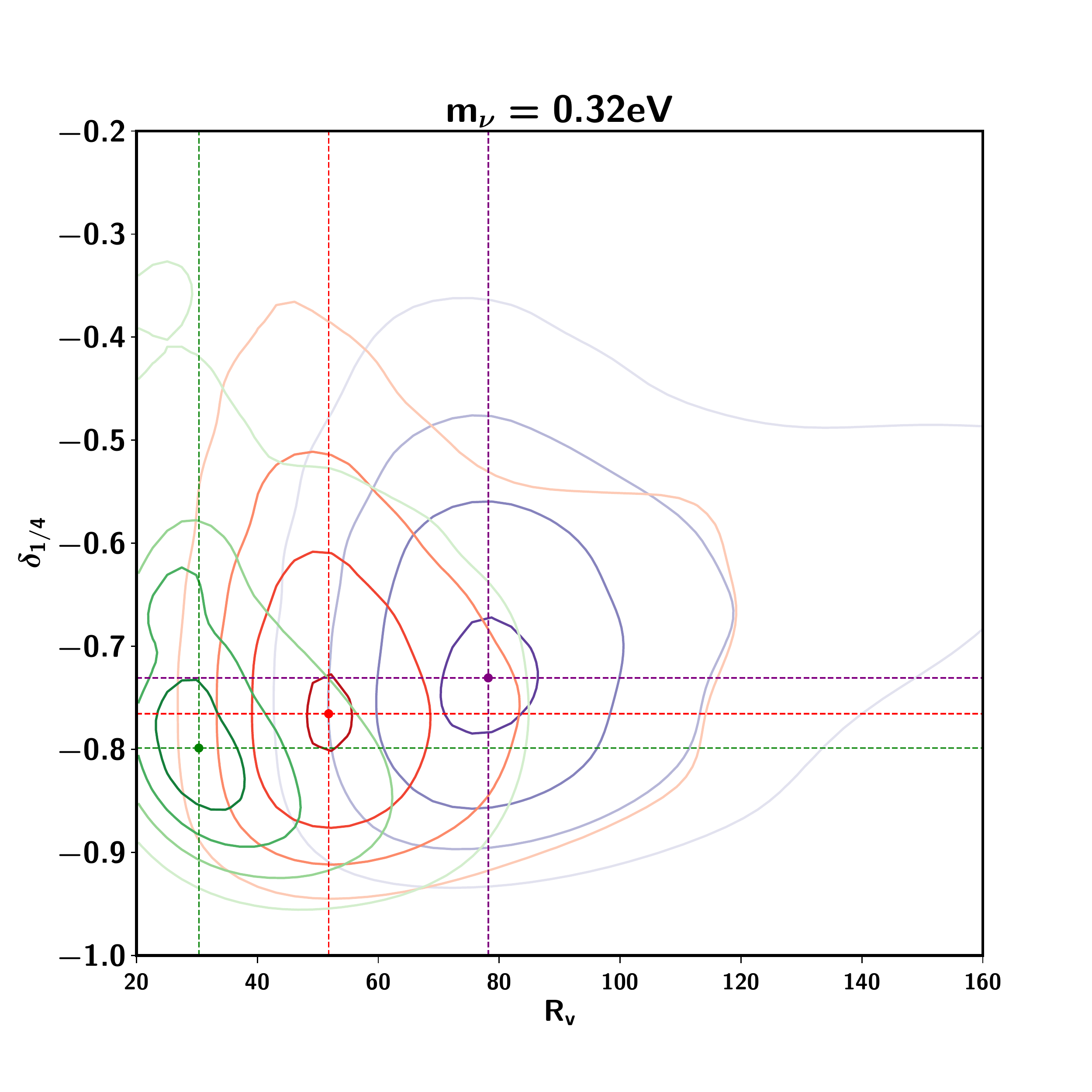}
\includegraphics[width=.4\columnwidth]{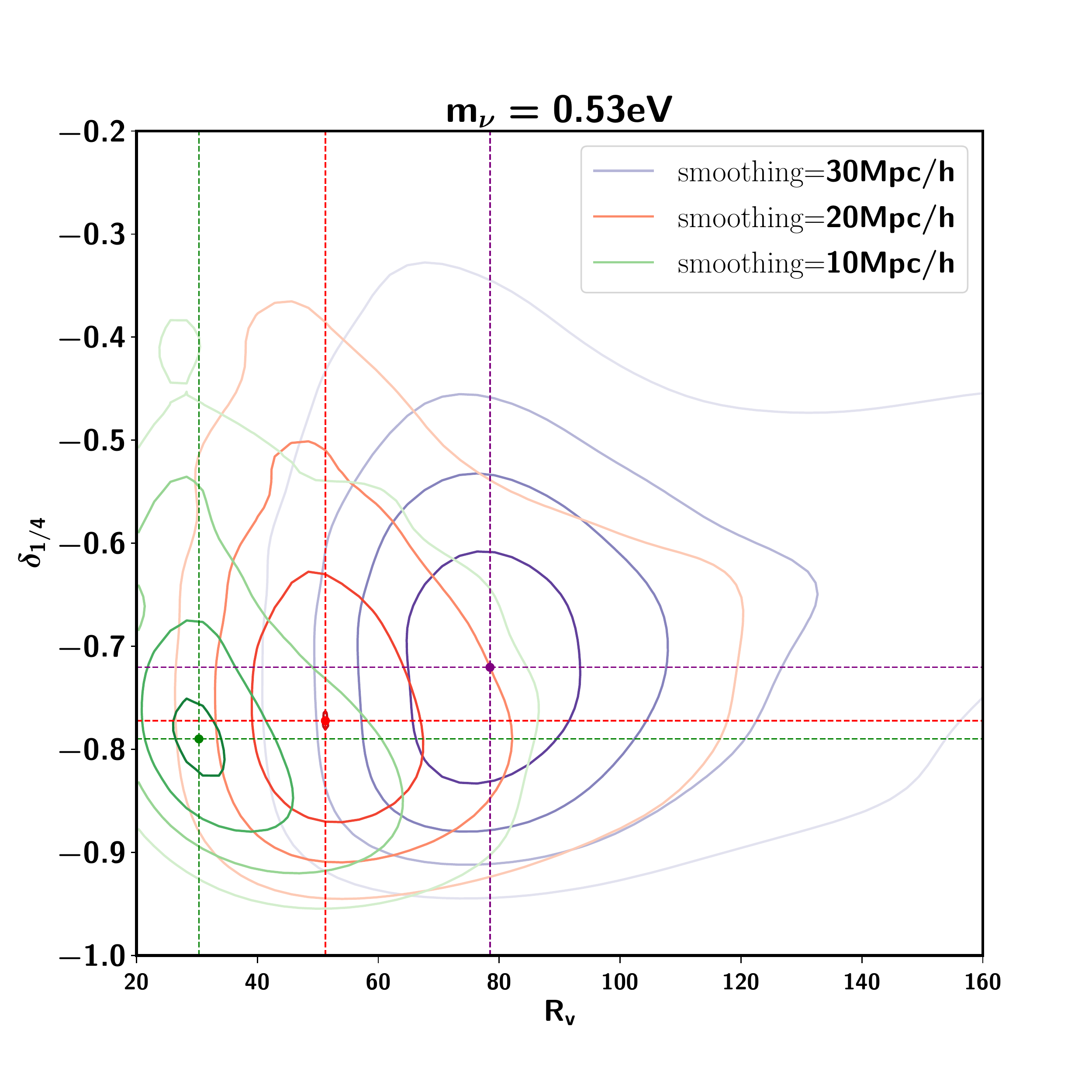}
\end{center}
\caption{Void distribution in the radius / central density plane $(R_{\rm v}, \delta_{1/4})$ for different smoothing scales, each panel representing the different massive neutrino fields in the DEMNUni simulations.}
\label{fig:central_dens_radii}
\end{figure}

Void finder parameters will impact the void properties and, crucially, the analysis itself. 
For example, in~\cite{Y1_ISW} and~\cite{Y3_ISW} using the same 2D void definition presented here, the authors chose a large gaussian smoothing parameter ($50 h^{-1}$ Mpc) that forces the detection of large underdense region, {\it supervoids}, with radius of the order of $R_{\rm v}\sim100 h^{-1}$ Mpc, for which a tension with $\Lambda$CDM without massive neutrinos simulations have been observed~\cite{granett2008,cai}. However, in the case of void signal in the CMB lensing map, the objects showing a stronger signal are medium-size voids, $R_{\rm v}\sim 40-80 \, h^{-1}$Mpc~\cite{vielzeuf2019}. 
As the aim of this work is to 
evaluate the impact of massive neutrinos in such signals, we have used three different smoothing prescriptions: 10, 20, 30 $h^{-1}$Mpc. 
In order to illustrate the effect of different smoothing, we show in Figure~\ref{fig:central_dens_radii} the distribution in central density ($\delta_{1/4}$) versus the void radius $R_{\rm v}$ computed for the void catalogues with different smoothing parameters, for the different neutrino masses prescriptions. Eventhough the effect of massive neutrino is difficult to detect, in the Figure we can see that increasing the smoothing scale of the void finder will result, on average, in voids smoother and larger. Namely, the most numerous population of voids will have radius of $R_{\rm v}\sim 30 h^{-1}$Mpc and a central density of $\delta_{1/4}\sim -0.8$ when the smoothing scale of the void finder is $10 h^{-1}$Mpc; with a smoothing scale of $30h^{-1}$Mpc we identify more voids with $R_{\rm v}\sim 75 h^{-1}$Mpc and $\delta_{1/4}\sim -0.7$.

Cosmic voids can be separated in different subgroups, in particular~\cite{sheth2004} presented two different scenarios at the origin of void formation:
\begin{itemize}
    \item \textit{voids-in-voids}: generally large underdensity, localized in an underdensed environment that are usually showing a negative relative bias with respect to the matter field~\cite{hamausbias}; 
    \item \textit{voids-in-clouds}: density minimum residing in larger overdensities, these voids are usually smaller objects compared to the previous ones.
\end{itemize}
Due to the local environment in which their reside, intrinsic properties of these two classes differ; in particular, their clustering properties will differ as in~\cite{zhao2016}, where the power spectrum of small voids ($R_{\rm v}<12h^{-1}$Mpc) has been shown to be correlated to the matter field, while large voids showed anti-correlation signal, {\it i.e.} small voids tend to reside in denser regions while larger voids will be identified in less populated regions. 
Therefore, we should expect differences in the correlation profile due to the environment; we check weather massive neutrinos could also affect the correlation signal differently, depending on the void population considered.
 Differentiating these populations is not straightforward, e.g. in~\cite{sheth2004} the authors used the void radius to separate the two type of voids, defining a \textit{characteristic void size} evolving with cosmic time. Moreover, as mentioned before, the presence of massive neutrino will smooth the density field up to a given scale, we expect effects due to their presence to depend both on the radius of the void considered and on their redshift.

\subsection{The void catalogue analysis}\label{sec:void_cat}

First, we investigate the properties of the void catalogues in order to understand the effect of massive neutrinos on voids.
In this section, we look at the the signature of massive neutrinos for different cosmic void parameters: i) the total void number, ii) void size function and iii) void density profile. The aim is to quantify how the void population changes when we add massive neutrino to the cosmic recipe.
We have then identified cosmic voids using the 2D void finder presented in Section~\ref{sec:void_finding} with three different smoothing parameters, $10 h^{-1}$Mpc, $20 h^{-1}$Mpc and $30 h^{-1}$Mpc using the DEMNUnii DM halo catalogues as tracers of the matter field, for the different $\Lambda$CDM cosmologies with and without massive neutrinos.

\subsubsection{Total void number and size}\label{sec:vsf}
Void statistics - and specifically the number of identified voids with a given size - has been exploited to constrain the cosmological model. In particular, the distribution of underdensed region with respect to their size have shown to be particularly sensitive to dynamical cosmological models~\citep{verza_2019}, and modified gravity models~\citep{barreira2015,Li2012,cai2015,Voivodic2017,Contarini2020}. Moreover, recent efforts measured the void abundance using a spherical model and the excursion principle, to perform cosmological analysis and set constraints on parameters~\cite{ronconi2017,contarini2019,contarini2022} and the fisrt cosmological constraint from void abundance have been recently performed using the BOSS DR12 observations~\cite{contarini2022boss}. Here, we investigate how the presence of massive neutrinos can impact void features and statistics, rather than modeling this effect.
Intuitively, one expects the presence of massive neutrinos to increase the total number of identified voids in the underlying matter field. In fact, massive neutrinos slowing down clustering process will in turn slow down void merging as well, resulting in a larger number of small voids. This has been observed in previous analysis~\cite{Kreisch2019,Schuster2019}, where the authors - using \texttt{VIDE}'s 3D voids identifier on the CDM field of the DEMNUni N-body simulation - have identified more voids while they were considering higher neutrino masses. However, \cite{Kreisch2019,Schuster2019} have shown that changing the matter tracer from CDM particles to the CDM halo field can in fact invert this trend. This can be explain by the fact that the massive neutrinos - as they are slowing down the clustering processes - modify as well the overall tracer density once we employ haloes as matter tracers (see Figure~\ref{fig:halo_mass_dist})\footnote{We note that this is not the case for CDM particles for which the overall density of particles will remain constant.}. This can directly impact the resulting void catalogues, since the sparsity of the tracer used to identify cosmic voids impacts both the void abundance and size~\cite{sutter2014}. Moreover, the direction of the change in the density of the tracers depends on the resolution of our simulation: higher resolution simulation will include smaller haloes, in even greater numbers once one accounts for massive neutrinos, resulting in a denser tracer sample. Lower resolution simulations, instead, will reach a turn-over point in the number of smaller haloes identified. In this work, we consider a void-tracer population with minimum mass $M_h\simeq 2.5\times 10^{12}h^{-1}M_\odot$, i.e. the minimum halo mass of the DEMNUni FoF catalogues, whose mass function, and therefore total density, have been shown to decrease due to the free streaming and growth suppression by massive neutrinos.

In Table~\ref{tab:voidnumber} we show the total number of voids identified for each smoothing parameter, and their ratio w.r.t. the massless neutrino case. First, we note that differences in the smoothing scale parameter of the void finder will impact the void properties: the smaller the smoothing scale, the larger the number of small voids (see~\cite{vielzeuf2019} and Figure~\ref{fig:central_dens_radii}). Then, in Table~\ref{tab:voidnumber} we measure a decrease in the total number of voids traced by haloes with mass $M_h>2.5\times 10^{12} h^{-1} M_\odot$ while the neutrino mass increases for the $10 h^{-1}$Mpc and $20 h^{-1}$Mpc smoothing parameter; this is consistent to what have been observed in~\citep{massara2015}. This effect decreases while we use higher smoothing scales, and it is, in fact, the opposite when we consider a $30 h^{-1}$Mpc smoothing scale. 

\begin{table}

 \centering   
\begin{tabular}{|l|r|r|r|}
\hline
\diaghead{\theadfont Diag ColumnmnHead II}%
{}{smoothing scale} & \text{$10 h^{-1}$ Mpc} & \text{$20 h^{-1}$ Mpc} & \text{$30 h^{-1}$ Mpc} \\ \hline
\text{$n_{\Lambda{\rm CDM}} $}                                                                        & 144,594             & 68,221              & 30,055              \\ \hline
\text{$n_{\Lambda{\rm CDM} + m_\nu=0.16{\rm eV}}$}                                                          & 129,957             & 65,563              & 30,767              \\ \hline
\text{$n_{\Lambda{\rm CDM} + m_\nu=0.32{\rm eV}}$}                                                          & 114,046             & 61,945             & 30,986              \\ \hline
\text{$n_{\Lambda{\rm CDM} + m_\nu=0.53{\rm eV}}$ }                                                                  & 98,658              & 58,016              & 30,814              \\ \hline
\text{$n_{\Lambda{\rm CDM} + m_\nu=0.16{\rm eV}}/n_{\Lambda{\rm CDM}}$}                                      & 0.90               & 0.96      & 1.02               \\ \hline
\text{$n_{\Lambda{\rm CDM} + m_\nu=0.32{\rm eV}}/n_{\Lambda{\rm CDM}}$}                                      & 0.80               & 0.91      & 1.03               \\ \hline
\text{$n_{\Lambda{\rm CDM} + m_\nu=0.53 {\rm eV}}/n_{\Lambda{\rm CDM}}$}
& 0.68               & 0.85               & 1.02              \\ \hline
\end{tabular}
\caption{Total number of cosmic voids (and ratio w.r.t. the massless neutrino $\Lambda$CDM case, bottom rows) traced by haloes with mass $M_h > 2.5\times 10^{12} h^{-1} M_\odot$, found in the DEMNUni simulations for different void finder smoothing scales and different  neutrino masses.}\label{tab:voidnumber}

\end{table}

\begin{figure*}[htbp]

\centering

\includegraphics[width=\columnwidth]{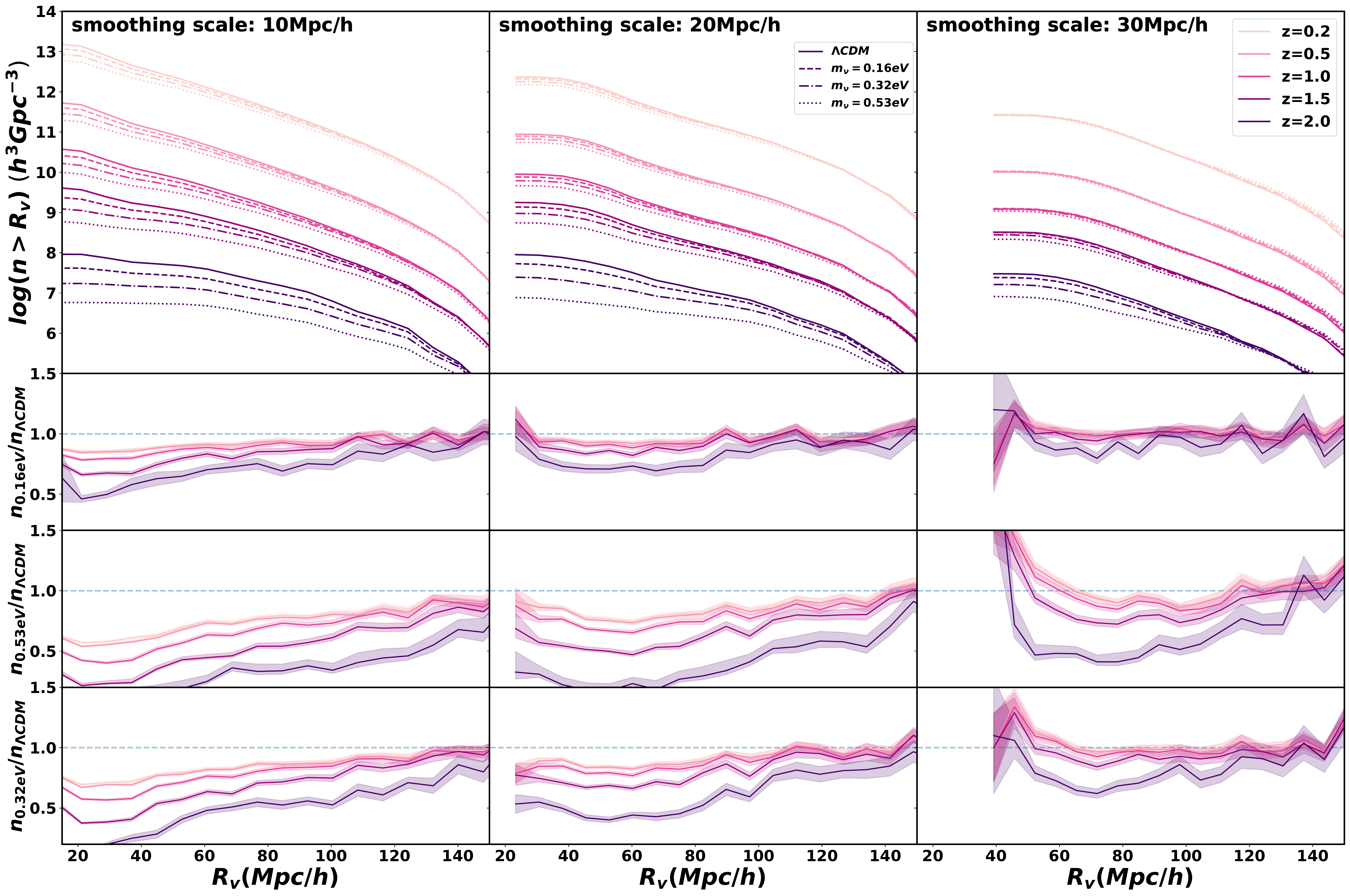}

\caption{{\it Top panel}: abundances of 2D voids at different redshifts, as a function of void radius in Mpc/$h$, for massless (solid line) and massive neutrino (dashed line) simulations. From left to right,  the different smoothing scales considered. {\it Bottom panels}: ratio of the void abundances in neutrino cosmologies w.r.t. the massless neutrinos $\Lambda$CDM case.} 
\label{fig:radius_dist}
\end{figure*}

The top panels of Figure~\ref{fig:radius_dist} show the 2D void abundances as a function of void radius for both massless neutrinos $\Lambda$CDM and $\Lambda{\rm CDM} + m_\nu$ simulations in various redshift bins, while the bottom panels show the ratio with respect to the massless neutrino cosmology. Within our definition of 2D voids, DM halo tracers show a drop in the number of small voids ($R_{\rm v}<50 h^{-1}$Mpc) if the neutrino are massive particles, and that this effect is even more pronounced for higher redshifts, which is consistent with the fact that at higher redshifts the range of scales affected by massive neutrino is larger than at lower ones (cf. Figure~\ref{fig:FS}). Moreover, similarly to Table~\ref{tab:voidnumber}, the choice of the smoothing scale parameter of the finder is related on how massive neutrinos are affecting the void size function. In fact, as we increase the smoothing parameter the effect of massive neutrino in the size function decreases: this effect could be explained since, as we increase the smoothing scale the finder tends to merge small voids into larger structures, and thus shifting the size function towards large voids, i.e. towards scales where neutrinos become non-relativistic, fall in potential wells, hence avoiding underdensed regions.

\subsubsection{Void density profiles}\label{sec:dens}

\begin{figure*}
\centering

\includegraphics[width=1\columnwidth]{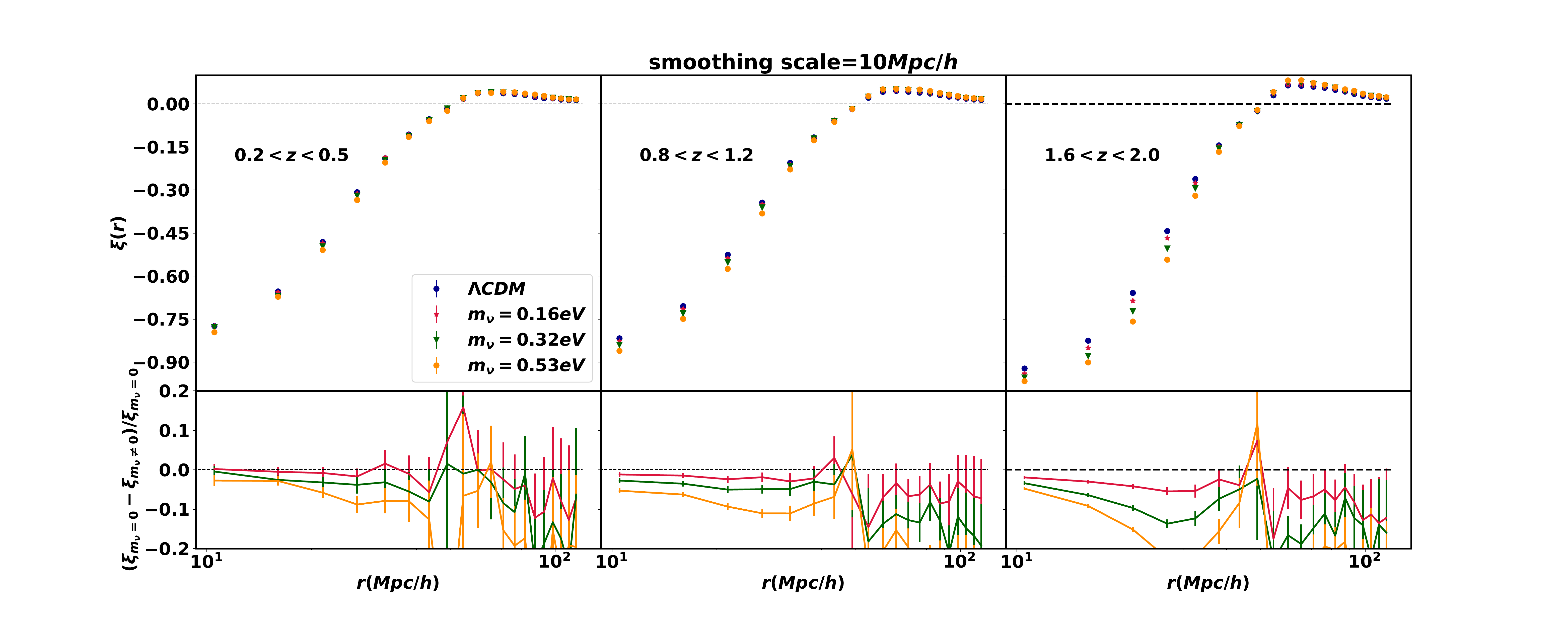}
\includegraphics[width=1\columnwidth]{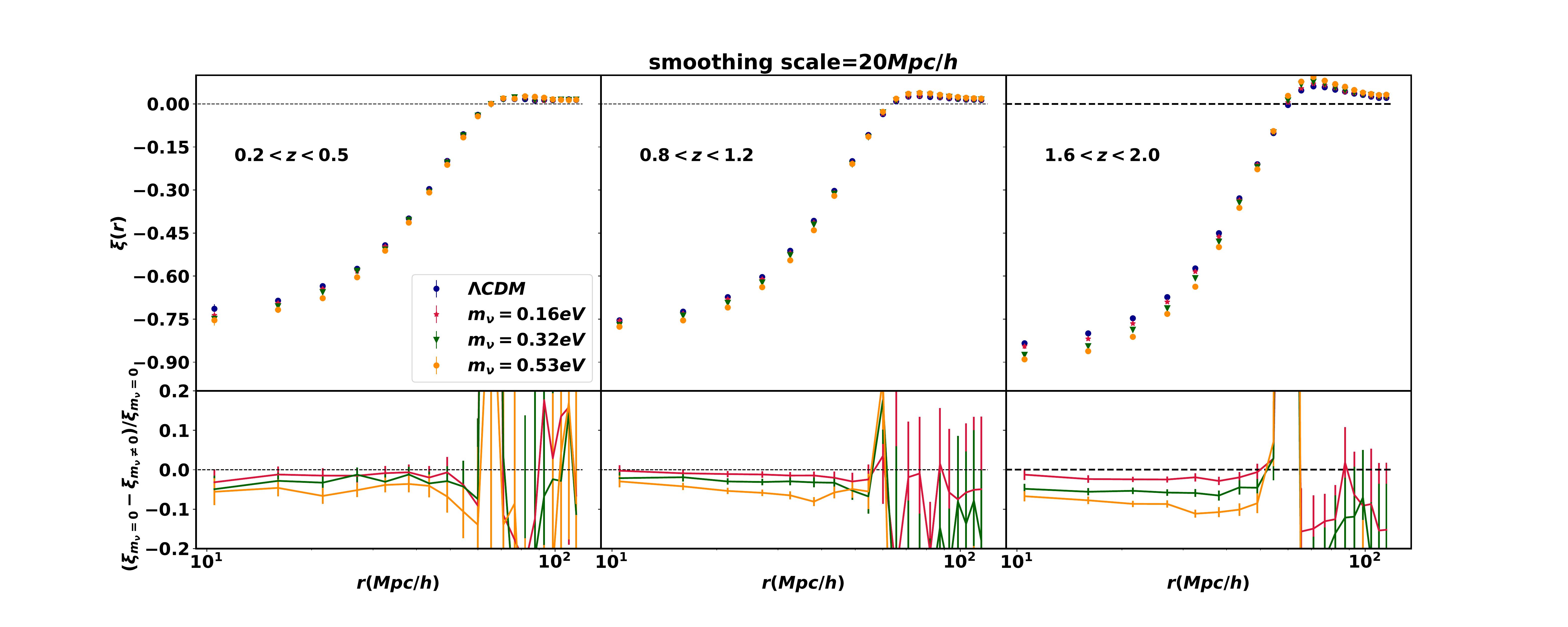}
\includegraphics[width=1\columnwidth]{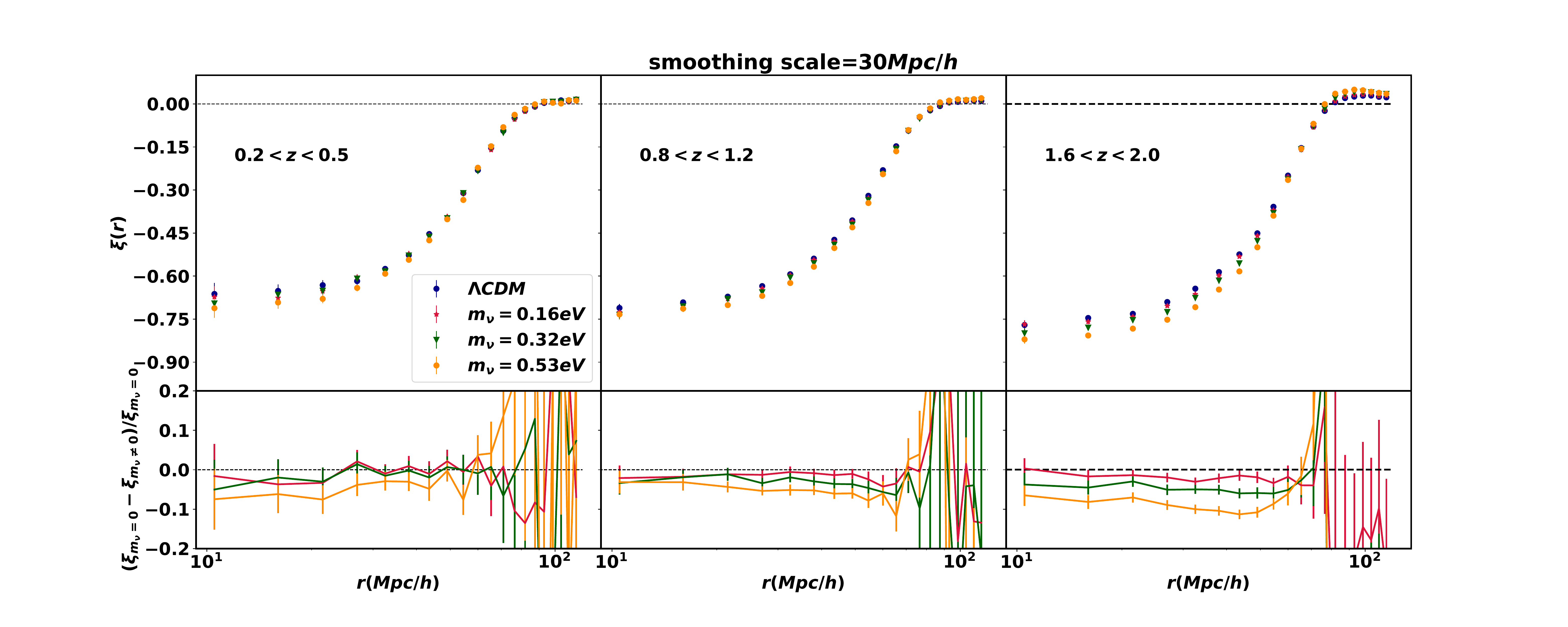}
\caption{Void-halo two point correlation functions, 2PCF. As in the void density profiles, each row shows a different smoothing scale, while each column a different redshift bin. The bottom panels in each box is the ratio between the massive neutrino 2PCF w.r.t. the massless neutrinos $\Lambda$CDM case.}
\label{fig:dens_prof_20}
\end{figure*}

The general density profile of cosmic voids has already been studied in several works and different models have been proposed in the literature~\citep{Ricciardelli2014,Hamaus2014,barreira2015}. However, these studies have highlighted the complexity of finding a general definition for this profile, due to the fact that it depends on the void definition itself (e.g. the choice of tracers of the matter field, void finder, smoothing scales, etc.). Nevertheless, in all these studies, cosmic voids can be described as underdensed regions at the void centre, surrounded by a more or less pronounced positive density shell at the void's edge: the so-called compensation wall. This wall is associated to filaments, while the depth of the central region and the size of the compensation wall will depend on the size of the considered objects. In this work, we will quantify how the presence of massive neutrinos can affect the density profile of the voids in the halo field. In fact, the density profile of cosmic voids, due to their scales, have shown to be  particularly sensitive to massive neutrino. In particular, from what have been observed in~\cite{Zhang2019} in simulations, the halo mass function inside voids is more affected by the presence of massive neutrinos than in other regions in the sky. This implies that the clustering process inside underdense regions, and consequently the void density profile, will be different while considering massive neutrinos in the cosmic budget.
In~\cite{massara2015}, by looking at the CDM density profile of 3D cosmic voids of a given size, they showed how the presence of massive neutrino will smooth the density profile by decreasing the size of the compensation wall and by making the void less empty at the void centre, and that this effect is more significant at low redshifts than higher ones. These effects are direct consequences of the slowing-down of clustering due to massive neutrinos.
However, it is also interesting to note that in~\cite{Contarini2020} the authors have shown that effects on the density profile of 3D voids in modified gravity models tend to be cancelled out when neutrinos are massive.

In the context of this work, our void definition allows us to identify voids much larger than the 3D voids studied in previous works, so that it becomes important to investigate the density profile of our 2D voids. In particular, changes in the underlying density profile of the objects are directly related to their imprint signals on the CMB convergence map. Thus, analysing the actual profile of our voids could give us insights to characterize their imprints in the CMB lensing map (see Section~\ref{sec:CMBxvoids}). 

The density profile of cosmic voids can be defined as the number of tracers at a given angular distance from the void centre, compared to the mean distribution of tracers at redshift $z$. We can analogously relate it to the void-halo two point cross-correlation function (2PCF), which refers to the measurement of pairs void/halo at different angular separations (see Eq.~4 in~\cite{Khoraminezhad2021}). Therefore, to estimate the density profile of our cosmic voids in the halo field, we have measured the void-halo 2PCF using the publicly available \texttt{GUNDAM} toolkit~\cite{gundam}. The \texttt{GUNDAM} pipeline measures the 2PCF $\xi_{ij}(r)$ directly in the ligthcone, using the Davis-Peebles estimator~\citep{DP}:
\begin{equation}
    \xi_{ij}(r)=D_iD_j(r)/D_iR(r)-1,
\end{equation}
where $D_{i}D_{j}(r)$ and $D_iR(r)$ represent the count of pairs at the distance $r$ between the object $i$ and the object $j$ or the random distribution of point $R$, respectively. 
Note that, in order to be in line with future experiments, we computed the density profile of our voids using DM haloes as tracers of the density, meaning that the resulting profile will be also related to the void size function of Figure~\ref{fig:radius_dist}.

In Figure~\ref{fig:dens_prof_20} we show the measured density profile for the void catalogue, divided in three redshift bins ($0.2<z<0.5$, $0.8<z<1.2$, $1.6<z<2.0$) for the three smoothing scales considered in this work (from top panel to bottom panel). We observe from the correlation function that once we consider larger smoothing scales, the identified voids tend to be smoother, with a central density less negative with respect to what is observed for smaller smoothing scales, which is consistent with what we saw in Figure~\ref{fig:central_dens_radii}.  
Moreover in all panels, we noticed that larger neutrino masses tend to make voids slightly deeper than in the massless neutrino $\Lambda$CDM case, which is an opposite trend to what has been observed in~\cite{massara2015}. However, the analysis of~\citep{massara2015} is using the non-observable dark matter particles as void-tracers, while in this work we use halo as void-tracers. It is well known that voids identified in the CDM field are and behave differently from voids identified in halo/galaxy distributions, which are biased tracers of the underlying dark matter field. In this respect, the profile amplitude of the halo-traced voids becomes deeper as the halo-bias increases with the neutrino mass~\cite{Marulli_Carbone_2011}, contrary to what we will observe in the next Section for the void profile measured in the lensing-convergence field, which decreases as massive neutrinos smooth the CDM density perturbations below their free-streaming scale. Since more biased objects are also sparser, the observed void profiles are thus in agreement with the decrease in the number density of gravitationally bound haloes with a mass greater than the minimum mass of the mock catalogues (see Section~\ref{sec:hmf}). Therefore, the more the neutrinos are massive, the more the voids will be devoid of massive objects. Similarly to what happens with the halo mass function, the effect of massive neutrino is greater at high redshifts than at low ones, the latter being also consistent with the fact that, at higher redshifts, larger scales will be affected by massive neutrinos (cf. Figure~\ref{fig:FS}) and more biased haloes will become even rarer. We stress that the density profile presented here is computed using DM haloes, i.e. biased objects as tracers of the matter field, as they are more realistic than the CDM particles which can be measured only in simulations.

\section{Voids CMB lensing cross-correlations}\label{sec:CMBxvoids}

Similarly to the correlation that have been observed between the CMB radiation and the overdensities identified in the foreground matter field \cite{baxter2015,Madhavacheril2015,planck2016,geach2017,baxter2018}, cosmic voids show, due to their underdensed nature, an anti-correlation (or a de-magnification) imprint in the lensing signal of the CMB. We will take advantage of the so-called stacking methodology \cite{krause2013,davies2018,higuchi2013} to reach high accuracy and a detectable signal-to-noise level, as the signal detected at a single void position can be noise-dominated \cite{krause2013}. This correlation between the CMB and voids has already been measured both in simulated and real catalogues \cite{cai,vielzeuf2019,Raghunathan2019}, and recent analysis have also shown a moderate discrepancy between observations and massless neutrinos $\Lambda$CDM simulations ($\sim 2-3\sigma$ lower signal in the observed data) \cite{hang2021,kovacs2022}. Furthermore, if confirmed, such a discrepancy could be related to the other tensions such as the correlation of super-voids and the CMB/ISW signal in simulated massless neutrinos $\Lambda$CDM mocks and observations \cite{granett2008,cai,Y1_ISW,Y3_ISW}.  
Our aim is to verify if this correlation signal can be affected by the neutrino mass, and if the presence of massive neutrinos could change the correlation signal itself, in the same direction as recently suggested by the aforementioned tensions.
Beside that, it is also important to stress that one of the powerful feature of the lensing signal comes from the fact that it is not directly influenced by the bias of matter tracers, since it is directly related to the true matter distribution inside the identified voids. Furthermore, in terms of cosmological probes, CMB lensing correlations with foreground objects also presents advantages with respect to background galaxy lensing: indeed as mentioned before, the peak of the CMB lensing kernel is around $z\sim 1.5$ and offers the opportunity to explore a wider and higher range of redshifts with respect to the lensing of background galaxies (see Fig.1 of \cite{farbod2016} for a comparison of the different lensing kernels). These correlations also present a particular interest in the sense that they are not affected by the main systematic errors that have to be considered when one estimate the lensing of background galaxy (such as intrinsic alignments or shear bias). 

\subsection{Imprint in CMB lensing map and stacking methodology}\label{sec:stacking_meth}

We construct a new estimator for the projected void density, in line with the previous definition of CMB convergence (Eq.~\ref{eq:kappa}). In fact, by looking at the position of a single underdense regions aligned with the CMB lensing reconstructed map, we would expect a negative convergence or a "de-lensing" signal, and this signal will be directly related to the underlying matter field, and thus to the projected void density profile. The convergence signal can then be expressed as a function of the projected void density in the angular direction $\boldsymbol\theta$:
\begin{equation}
    \kappa_{\rm v}(\boldsymbol\theta)=\Sigma_{\rm v}(\boldsymbol\theta)/\Sigma_{\rm crit};
\end{equation}
with the critical surface density for voids being defined as:
\begin{equation}
    \Sigma_{\rm crit}=\frac{c^2}{4\pi G}\frac{f_{\rm CMB}}{f_{\rm voids}f_{\rm CMB-voids}} ,
 \end{equation}
where $f_{\rm CMB}$, $f_{\rm voids}$ and $f_{\rm CMB-voids}$ are respectively the angular diameter distances to the CMB, to the considered void and between the CMB and the void. $\Sigma_{\rm v}(\boldsymbol\theta)$ represents the projected underlying distribution of matter as a function of the void centre, different from the density profile presented in Section~\ref{sec:dens} where we measured the density profile of the voids using dark matter haloes, which are biased objects of the underlying matter field.

In the literature, different stacking methods \cite{nadathur2016,Raghunathan2019} have been proposed;
in the context of this work, we decided to smooth the CMB with a Gaussian kernel with a Full Width at Half Maximum (FWHM) of 1 degree that has shown to be a good compromise to optimise the detection level \cite{vielzeuf2019}. To estimate the stacked correlation profile, we applied the same methodology presented in \cite{vielzeuf2019}, i.e. after cutting patches of 5 times the void radius ($R_{\rm v}$) in the smoothed CMB \healpix{} maps centreed at the position of the voids in our catalogues, we re-scaled the patches in order to have regions of similar sizes, and stacked them pixel by pixel. Once the stacked image is computed, it is possible to reconstruct the averaged convergence profile of the voids by averaging the pixel value in concentric shells around the image centre. Furthermore, \cite{nadathur2017} has shown using simulations that these imprints change according to the void population. In fact, the lensing imprint is directly related to the true underlying gravitational potential, and the inner density of cosmic voids - which depends on size and voids definition \cite{Hamaus2014,dai2015} - will impact on the strength of this correlation signal.

In order to estimate the error in our measurements, similarly to \cite{vielzeuf2019}, we will generate $1,000$ random CMB lensing maps using the \texttt{synfast} and \texttt{anafast} module from \healpix{} package, with the same power spectrum as the original DEMNUni map presented in Figure~\ref{fig:cls_kcmb_dem}. We then can compute the covariance of the cross-correlation signal using the stacking method on the void catalogues and these CMB lensing maps. We do not consider additional noise due to observational systematics, since our goal is to disentangle the role of neutrinos in cross-correlations from a pure physical point of view; we leave a more realistic measurement for forthcoming analysis. Nevertherless, we also verified the importance of non-gaussian terms in the noise of the CMB lensing reconstruction by building the covariance applying 1000 random rotations to our void catalogue and stacked them. The resulting errors have shown to be consistent with the Gaussian noise realisations method described above.

\subsection{Measurement of the voids-CMB lensing cross-correlation signal in the DEMNUni simulations}

As mentioned before, one of the advantage of imprints of structures on the CMB lensing field resides in the fact that the resulting signal is directly dependant on the underlying matter field. Once one includes massive neutrino to the cosmological model, imprints of the different matter can be affected in two ways:
\begin{itemize}
    \item the lensing imprint due to neutrinos alone, which changes the overall lensing signal amplitude due to their mass. In fact, we are adding a non-relativistic component to the matter field, an additional particle that will enhance lensing to the background radiation. One expects this effect to occurs at scales larger than the free-streaming scale, i.e. at scales where neutrinos fall in potential wells.
    \item The change in the lensing signal due to the slowing-down of clustering of cold dark matter particles, caused by the presence of massive neutrinos.
\end{itemize}
In this section we will explore both effects for the different neutrino masses recipes.

\subsubsection{Neutrino contribution to the correlation signal}\label{sec:neutr_contr}

First, we investigate the contribution of massive neutrinos themselves to the correlation signals. In other words, we look at the correlation signals of the different elements (CDM and massive neutrinos) separately for the different parametrizations of the void finder. We have applied our stacking methodology presented in Section~\ref{sec:stacking_meth} to both lensing maps: the lensing signal of CDM-only and the CMB lensing map due to massive neutrino only, for the void catalogue identified in the DEMNUNi simulation with $m_\nu=0.53$eV.

\begin{figure}[htbp]
\begin{center}
\hspace*{-1cm}\includegraphics[keepaspectratio,height=20cm, width=20cm]{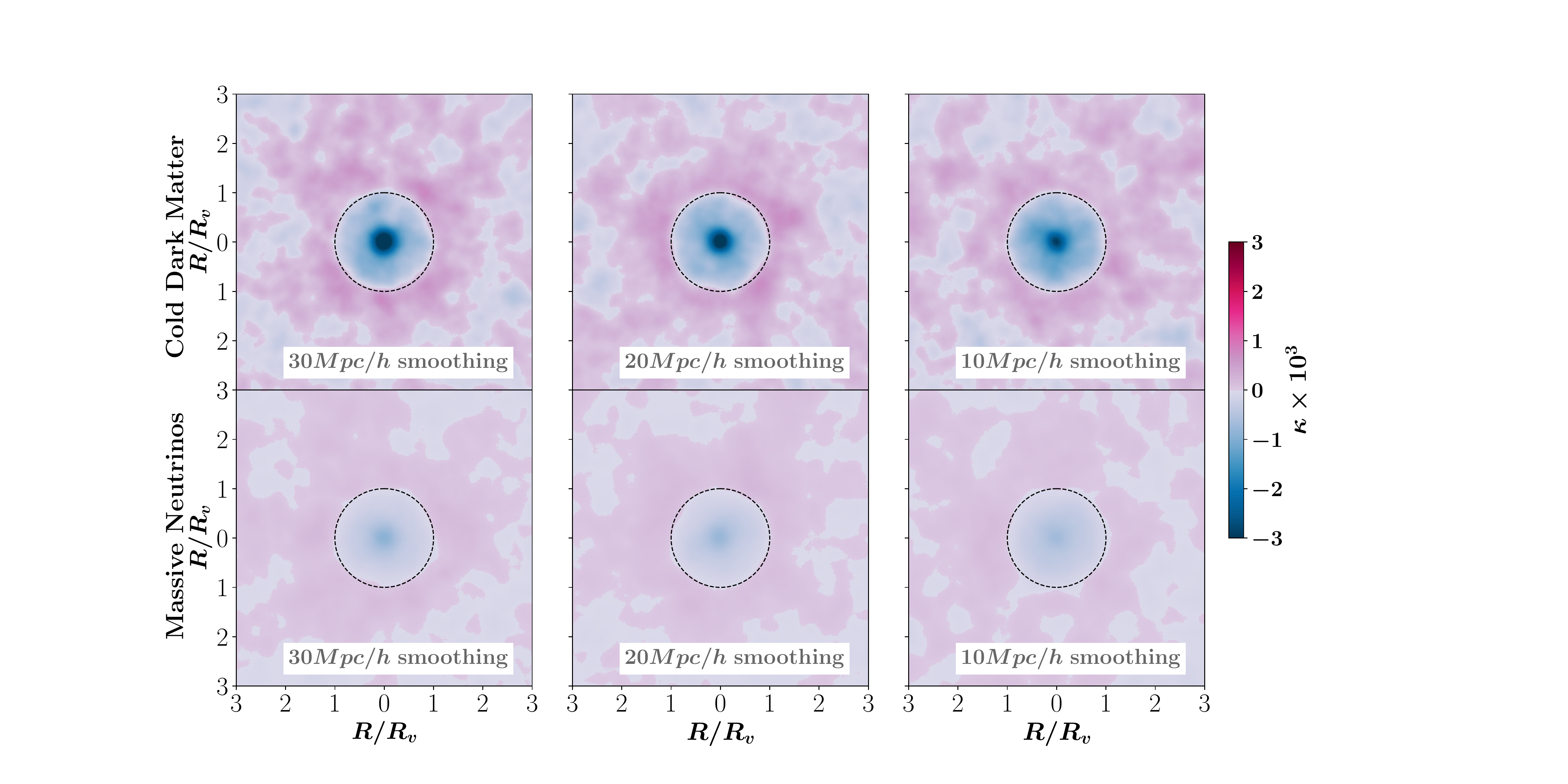}
\includegraphics[keepaspectratio,height=8.5cm, width=8.5cm]{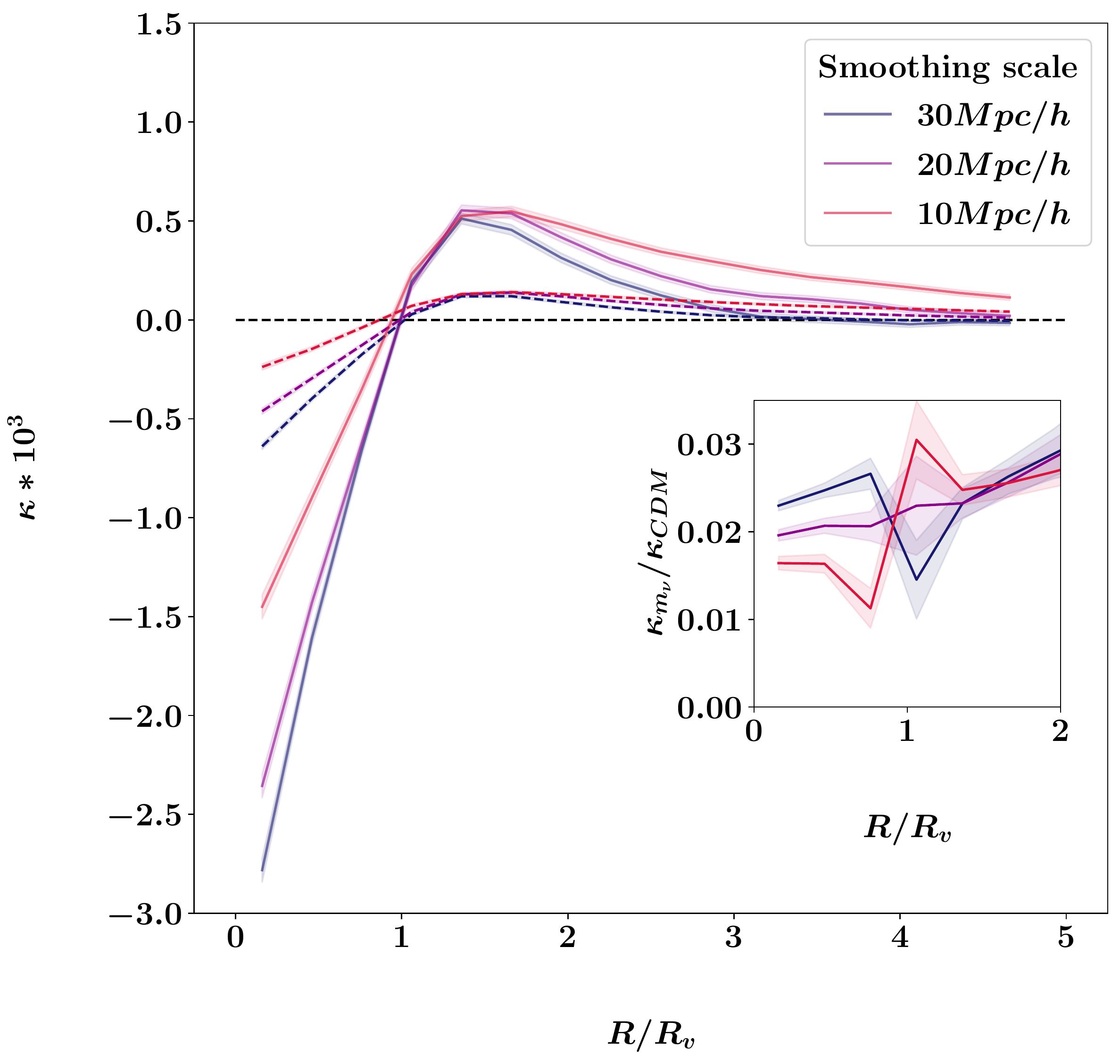}

\caption{{\it Top panel:} Stacked image of the imprint of the different void catalogues on CMB lensing, CDM-only component (top row) and massive neutrino only component with $m_\nu=0.53$ eV (bottom row). {\it Bottom panel:} Contribution of neutrino in the stacking imprint of cosmic voids (dash line) compared to the CDM contribution (solid line) for the three different smoothing scale considered. Note that the amplitude of the neutrino contribution on this figure has been amplified by a factor of 10. The shaded regions represent the error computed via 1,000 realisations of CMB lensing maps as explained in Section~\ref{sec:stacking_meth}.} 
\label{fig:neutrinocontrib_im}
\end{center}
\end{figure}

In Figure~\ref{fig:neutrinocontrib_im}, we show the stacked images of the two lensing imprints, CDM-only (top panels) and massive neutrinos-only (bottom panels) for the three smoothing scales considered (from left to right). As expected, we observe a negative imprint in the stacked lensing signal at the void positions for both the cold dark matter and neutrino field: voids are in fact underdense regions in both the fields. We note that, since neutrino density perturbations are much lower than the CDM ones, their imprint on the CMB lensing map is smaller than for CDM. In the bottom panel of Figure~\ref{fig:neutrinocontrib_im}, we show the corresponding lensing profiles as a function of the distance to the void center (normalized with the void radius $R_{\rm v}$) for the contribution of massive neutrinos (dashed lines) w.r.t. CDM (solid lines, and for the different smoothing scales considered (10, 20 and 30 $h^{-1}$Mpc). The neutrino imprint signal has been multiplied by a factor of 10 for visualisation in the same Figure as the cold dark matter one. The ratio of massive neutrinos to CDM as a function of distance to the void center ($\kappa_{m_\nu}/\kappa_{\rm CDM}$) is shown in the insight panel of Figure~\ref{fig:neutrinocontrib_im}. As expected, the neutrino-only contribution to the total void-CMB lensing cross-correlation represents few percents of the total signal. The figure also shows that as we increase the smoothing parameter of the void finder, the relative contributions of both neutrinos and CDM to the signal tends to increase. This is consistent with \cite{vielzeuf2019}, and can been seen as the consequence of a selection effect induced by the smoothing process in the void identification. Indeed, the smoothing kernel applied to the density field will force the algorithm to neglect structure with scales below the smoothing parameter. These structure will anyway lie inside our voids, thus including more nonlinear modes inside the voids, and these nonlinear fluctuations will boost the amplitude of the correlation signal making the voids deeper as the smoothing scale increases. 
We stress that this behaviour seems to be opposite to the one of the two point correlation function shown in Figure~\ref{fig:dens_prof_20}. However, this is only partially due to the difference between the density profiles measured in the halo-traced void field and CDM-traced void field, respectively, while the main effect comes from the smoothing length of our void-finder.
Moreover, while in Figure~\ref{fig:dens_prof_20} the profile of the halo-traced voids becomes slightly shallower as we increase the smoothing length, the void profile measured in the lensing-convergence field becomes deeper.
A similar trend can be observed also the neutrino-convergence void-profile, meaning that similarly to CDM, neutrinos seems to be less present in voids identified using large smoothing scales, and thus we can measure a stronger void-lensing signal if we increase the smoothing scale. 
However, if we look at the insight plot, we see that as we increase the smoothing scale, the ratio of the signals from massive neutrinos and CDM increases. This implies that, by increasing the smoothing scale, the abundance of massive neutrinos inside the voids decreases faster than the CDM abundance. This could be explained by a possible reduction of the massive neutrino abundance at small scales, w.r.t the CDM one, but higher resolution simulations would be required to confirm such results.

Moreover, as shown in Figure~\ref{fig:FS}, massive neutrinos will fall in potential wells of different sizes and this will be redshift dependent. In Appendix \ref{sec:appendix_A} we have tested such a behaviour by binning our sample in different radius bins and redshifts. Even though the redshift evolution of the separate signals of massive neutrino and CDM is difficult to observe, we could detect a variation of these two contributions once we consider different void sizes, suggesting, in agreement with theoretical predictions, a stronger presence of massive neutrinos in smaller voids.

\subsubsection{Cross-correlations CMBL and voids}
\begin{figure*}
    \centering
    \includegraphics[width=50mm]{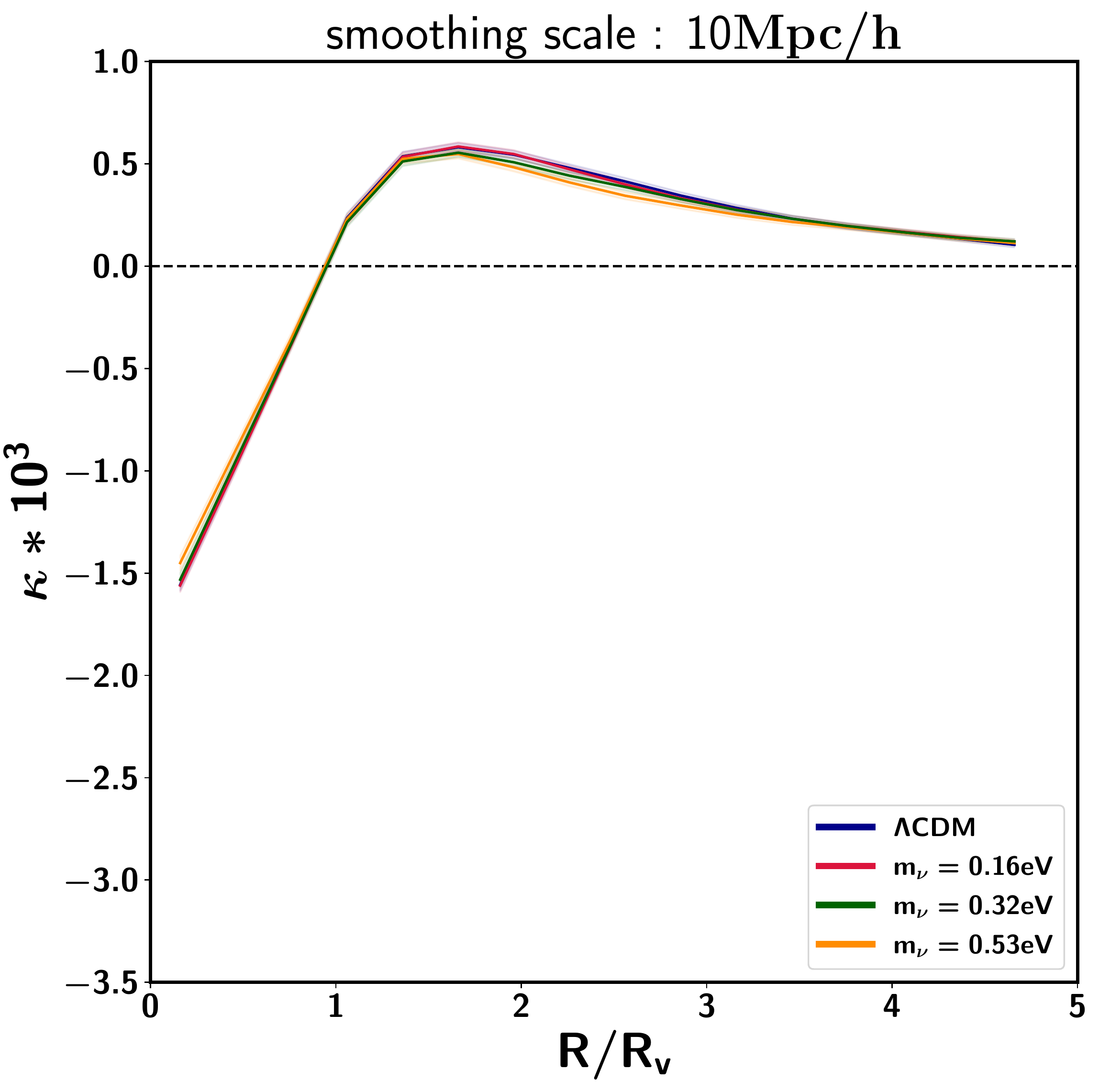}
    \includegraphics[width=50mm]{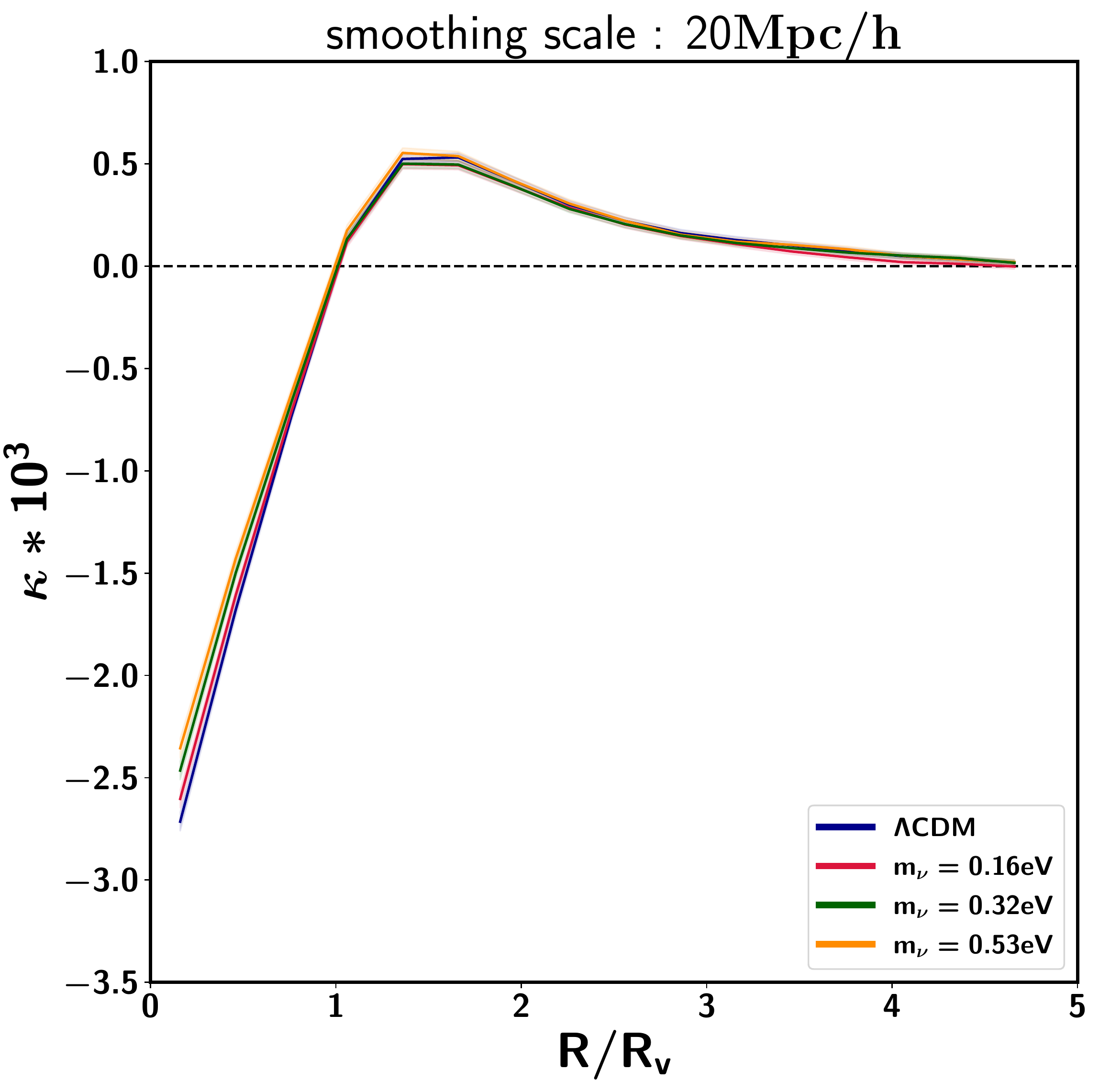}
    \includegraphics[width=50mm]{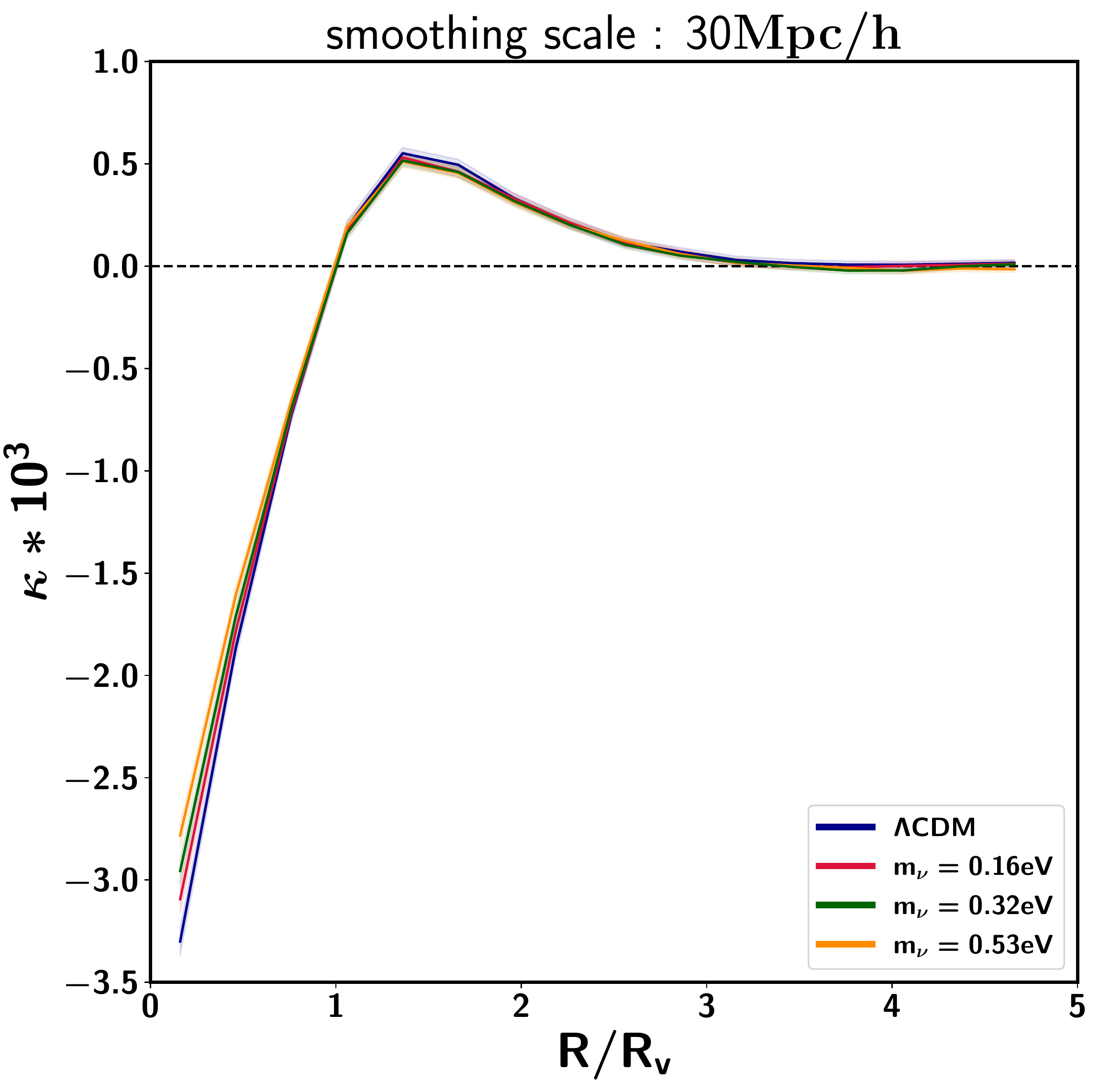}
    \caption{Imprint of cosmic voids for different massive neutrino cosmologies, combined CDM and neutrinos. From left to right, three smoothing scales: 10 $h^{-1}$Mpc, 20 $h^{-1}$Mpc, 30 $h^{-1}$Mpc .}
    \label{fig:FULL_SIGNAL}
\end{figure*}

\begin{figure}
    \centering
    \includegraphics[width=.5\columnwidth]{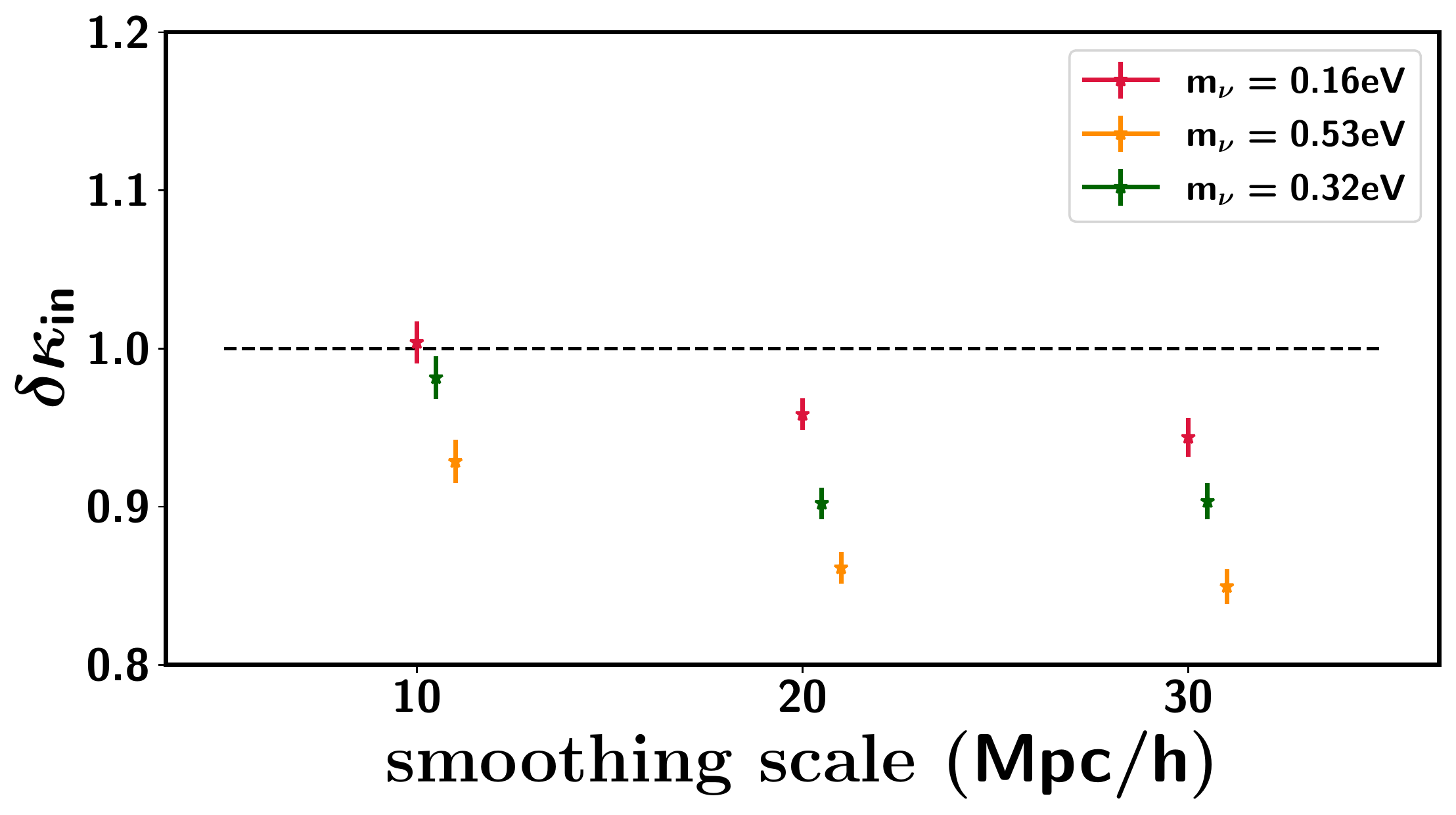}
    \caption{Sensitivity parameter $\delta\kappa_{in}$ - Eq.~\eqref{eq:sens_param} - for different massive neutrino masses as a function of the void finder smoothing scale.}
    \label{fig:FULL_sensiv}
\end{figure}

In this section, we analyse the full cross-correlation signals (from neutrino plus CDM particles) obtained for all voids catalogues and the different neutrino masses presented before. Figure~\ref{fig:FULL_SIGNAL} shows the correlation profile for the different void catalogues. Similarly to previous results, the strength of the lensing imprint on the CMB caused by cosmic voids will be larger as we increase the smoothing parameter in the void finder. Moreover, the differences in this signal due to massive neutrinos appear larger for larger smoothing scales. As expected, as we increase the neutrino mass, the lensing amplitude at the void center decreases, implying a slowing down of matter density perturbations caused by free-streaming neutrinos. While we can observe differences between the different cases inside the voids, moving further outside from the void centre all signal seems to converge. 
We want to quantify the sensitivity level of each measurement: to this purpose we consider the CDM+$\nu$ lensing-convergence within the void region where it varies the most ($R<R_{\rm v}/2$), and compute its ratio for different neutrino masses. In this way we define $\delta\kappa_{\rm in}$, i.e. a sensitivity parameter to the neutrino mass of the lensing-covergence void profile :
\begin{equation}\label{eq:sens_param}
    \delta\kappa_{in}=\frac{\sum_0^{r<R_{\rm v}/2}\kappa_{m_\nu=0.16{\rm eV},0.32 {\rm eV},0.53{\rm eV}}}{\sum_0^{r<R_{\rm v}/2}\kappa_{\Lambda {\rm CDM}}},
\end{equation}
where $\delta\kappa_{in}$ stands for the amplitude ratio of the signal with and without massive neutrinos in the inner region of the void ($R<R_{\rm v}/2$)\footnote{Note that this amplitude parameter is similar to the lensing amplitude parameter $A_\kappa=\kappa_{\rm obs}/\kappa_{\rm sims}$ used in the literature (see e.g. \cite{vielzeuf2019,kovacs2022}) to evaluate the agreement between the observed correlation signal, $\kappa_{\rm obs}$, and the one measured in $\Lambda$CDM simulations, $\kappa_{\rm sim}$.}. Figure~\ref{fig:FULL_sensiv} shows the sensitivity parameter of Equation~\ref{eq:sens_param} as a function of the smoothing scale of the void finder. As already observed in Figure~\ref{fig:FULL_SIGNAL}, the increase in the smoothing scale in the void finder results in a boost of the intensity of the cross-correlation signal amplitude, and this boost seems to be dependant on the mass of the neutrinos present in the simulations. In other words, as we increase the smoothing scale of the void finder, we measure a larger difference in the correlation signal of massive neutrino simulations with respect to the massless neutrino $\Lambda$CDM cosmology. The errorbars in Figure~\ref{fig:FULL_sensiv} have been estimated by propagating the errors of our stacking measurement described in Section~\ref{sec:stacking_meth}, thus not considering any extra systematic errors. 
The measure of this reduction in the lensing signal inside cosmic voids due to the presence of massive neutrinos is in particular interesting as it is consistent with the tensions in the recently observed lensing signal and massless neutrinos $\Lambda$CDM simulations \cite{hang2021,kovacs2022}. Namely, in both analysis voids have been identified using the 2D void finder described previously with a smoothing scale of $20 h^{-1}$Mpc and $10 h^{-1}$Mpc respectively, resulting in a observed signal of the correlation of cosmic voids with the Planck 2018 lensing convergence map \cite{planck2018} about $2\sigma$ lower than the one measured in massless neutrinos $\Lambda$CDM simulation without massive neutrinos. The direction of this tension is thus in line with the decrease of the lensing imprint of cosmic voids caused by the presence of massive neutrinos in our simulations.

\subparagraph{Void redshift evolution:}

\begin{figure*}

\centering

\includegraphics[width=.32\textwidth]{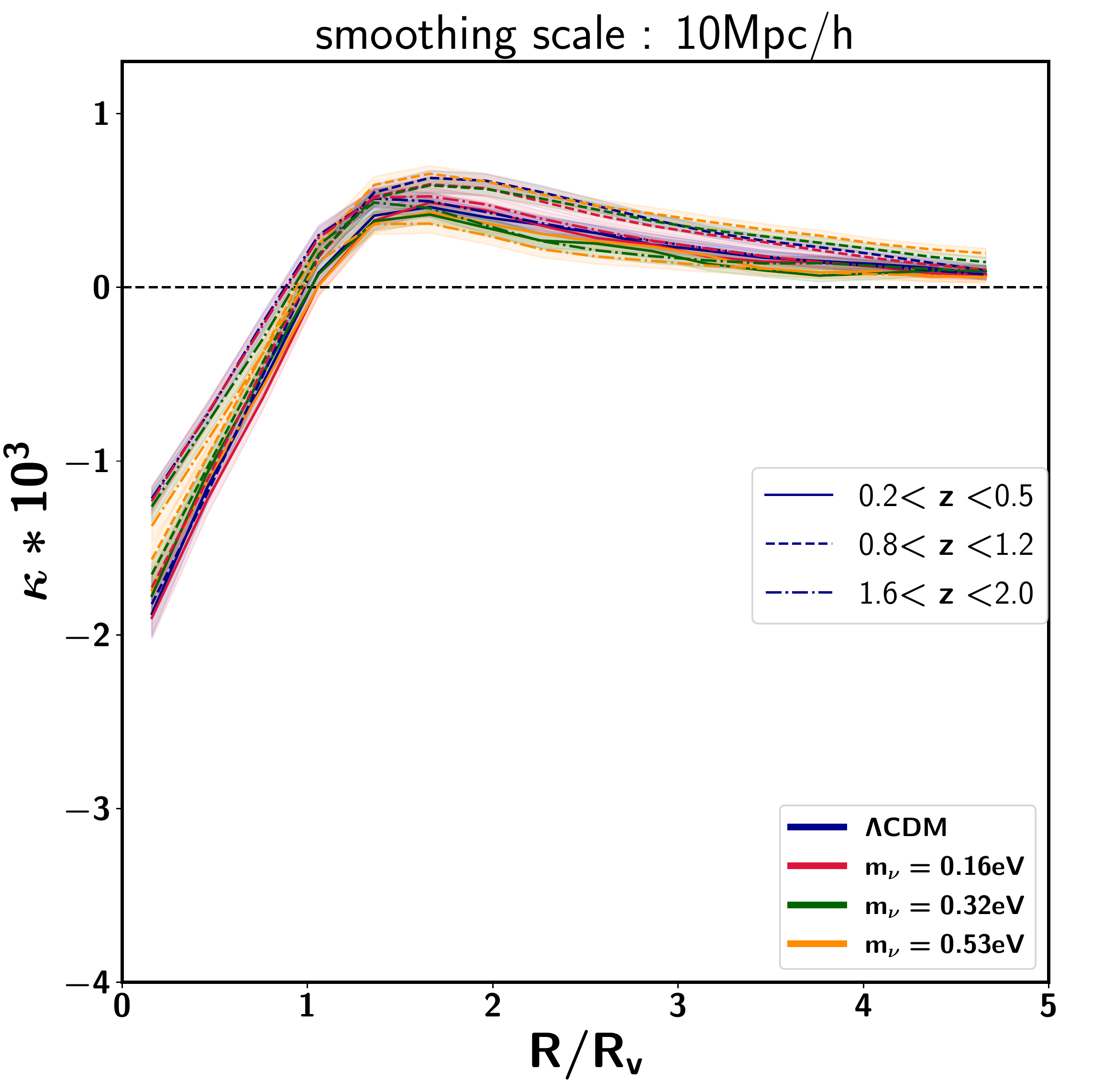}
\includegraphics[width=.32\textwidth]{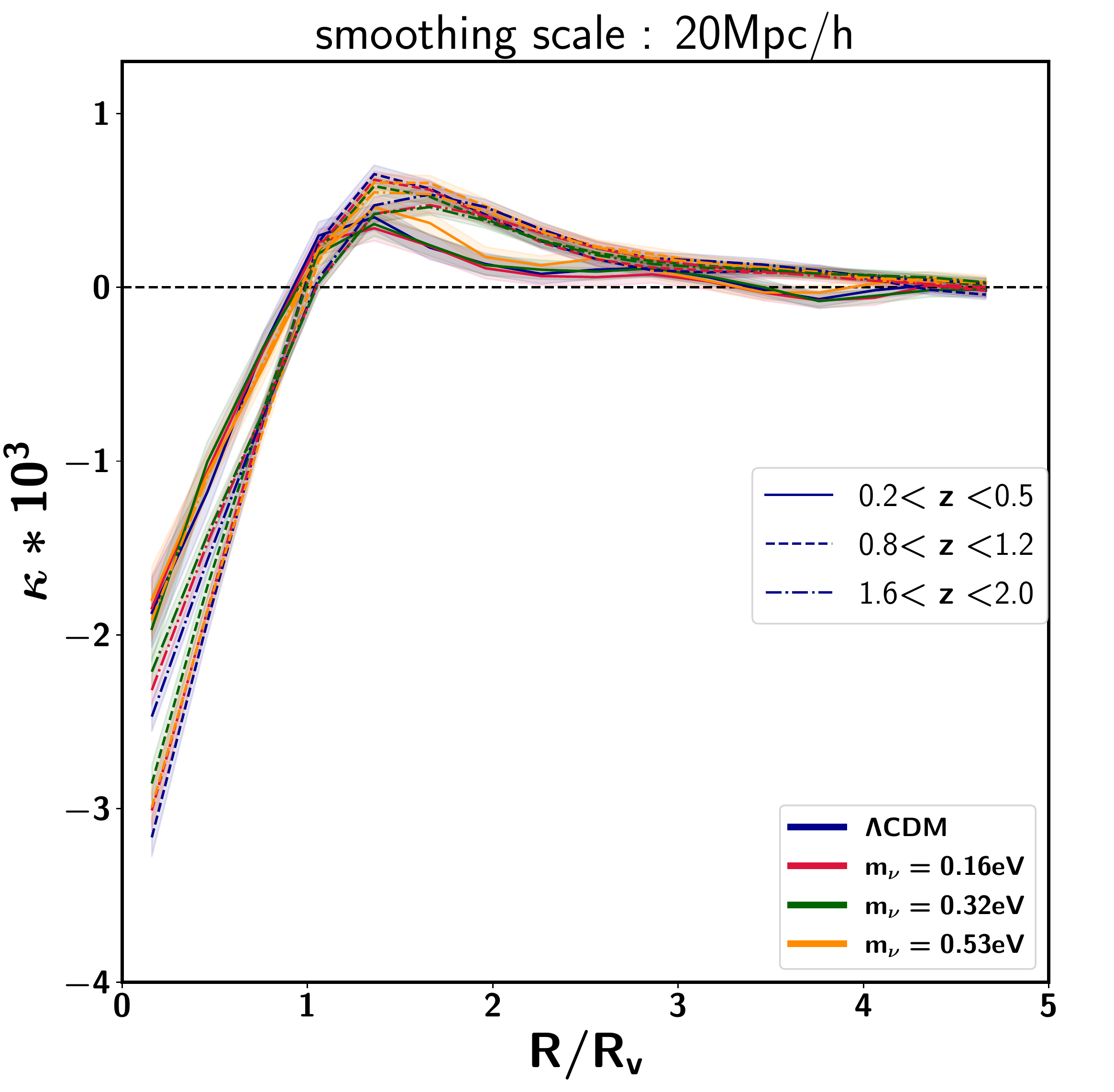}
\includegraphics[width=.32\textwidth]{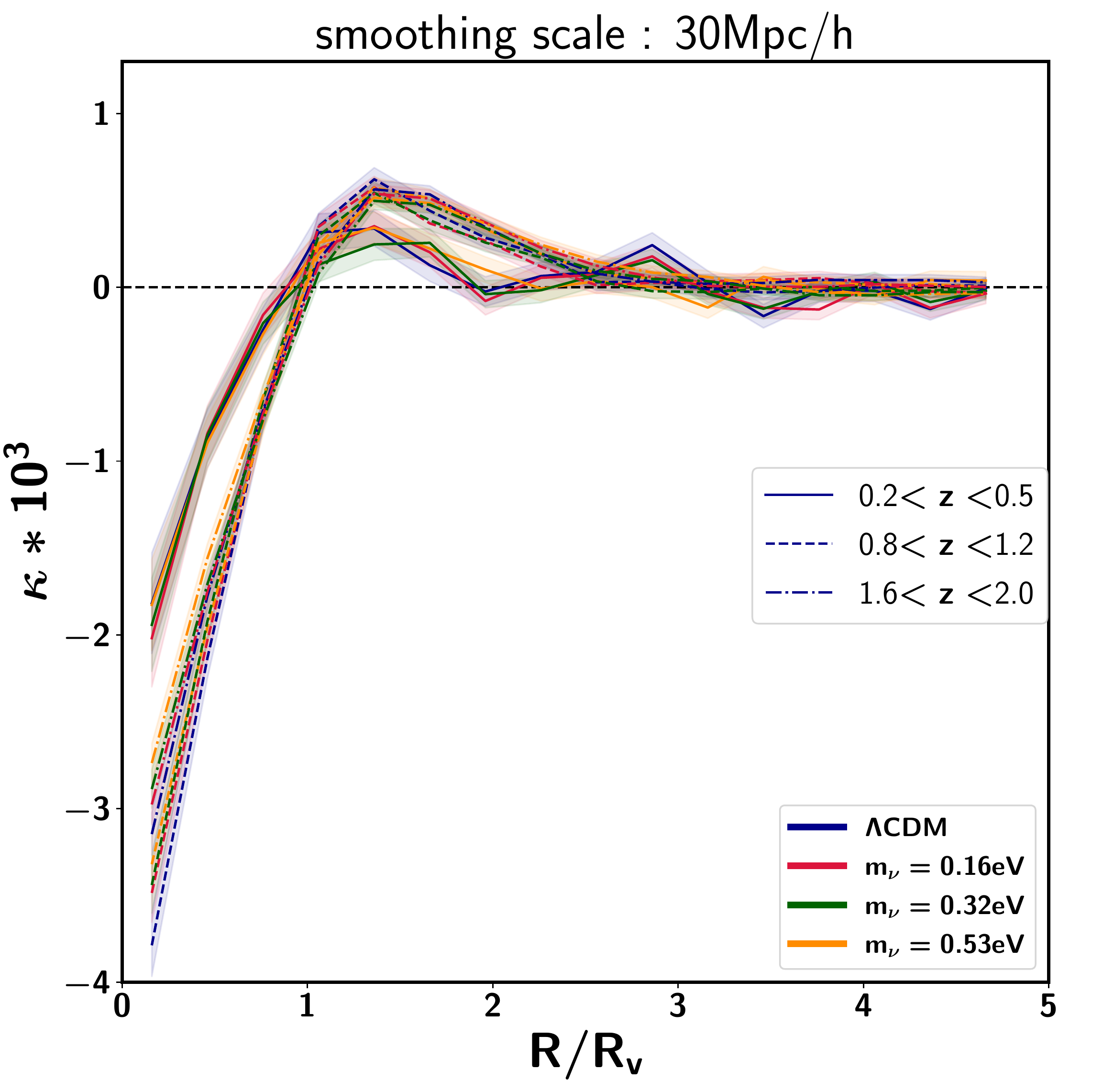}

\includegraphics[width=.32\textwidth]{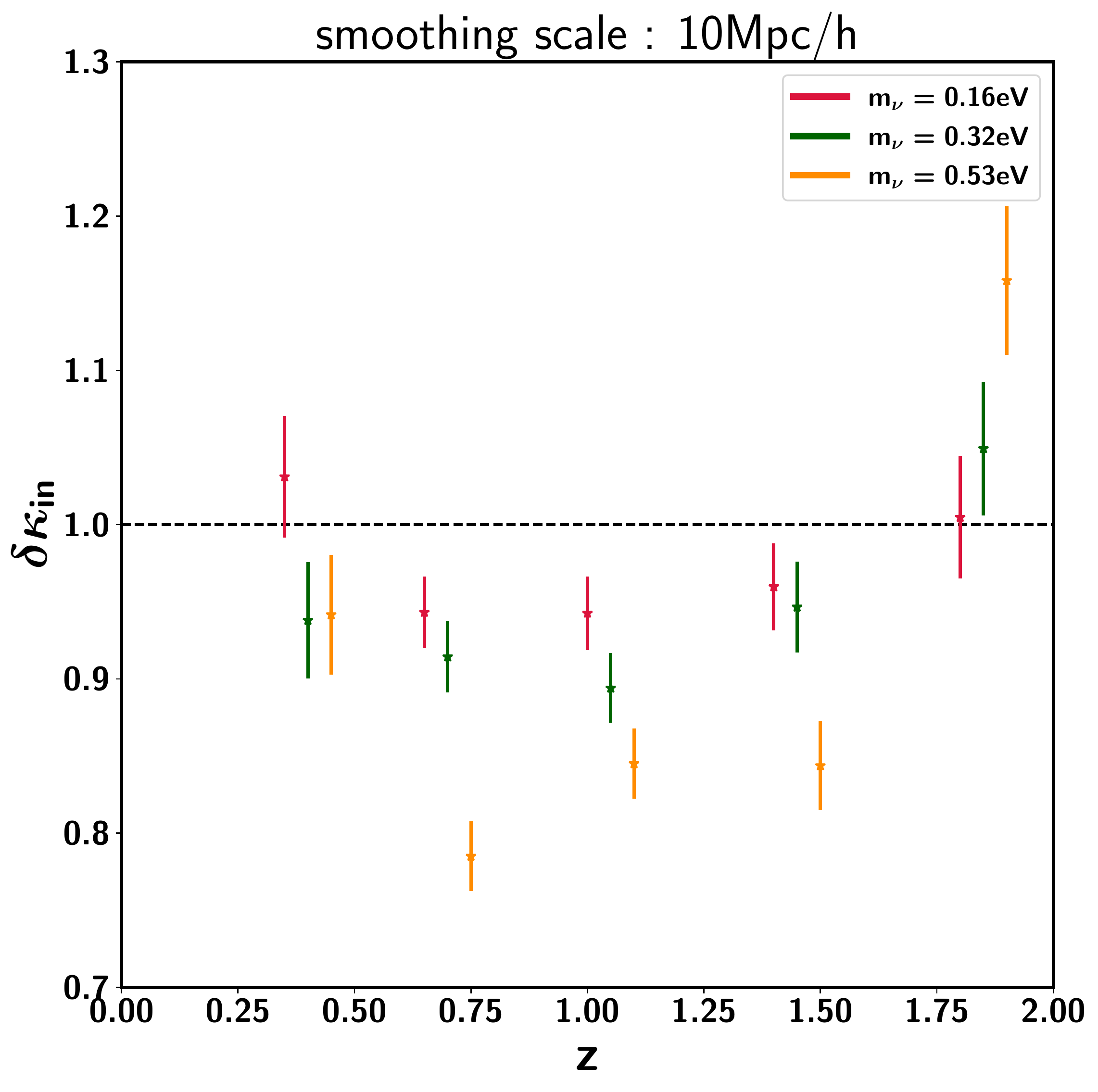}
\includegraphics[width=.32\textwidth]{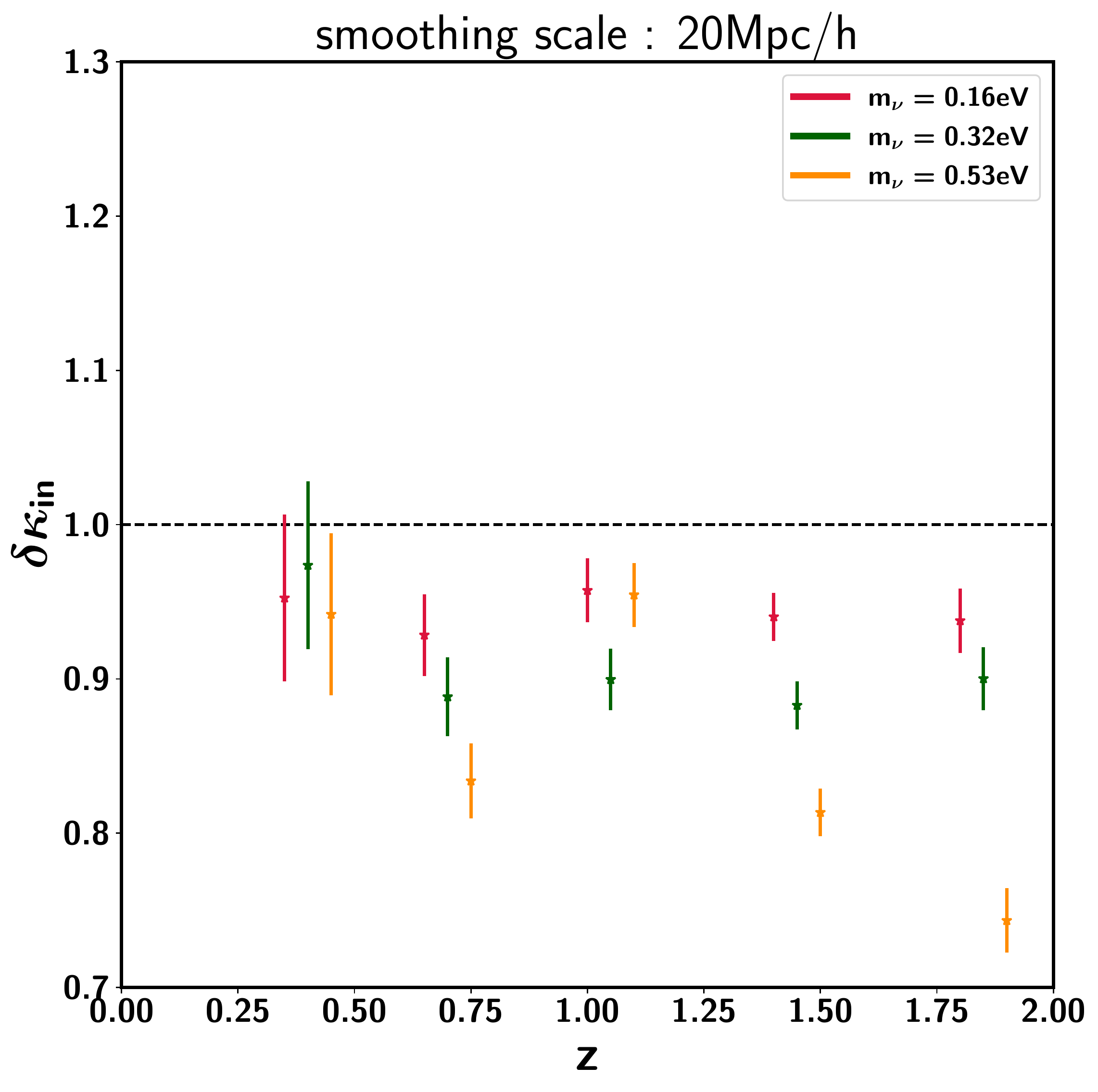}
\includegraphics[width=.32\textwidth]{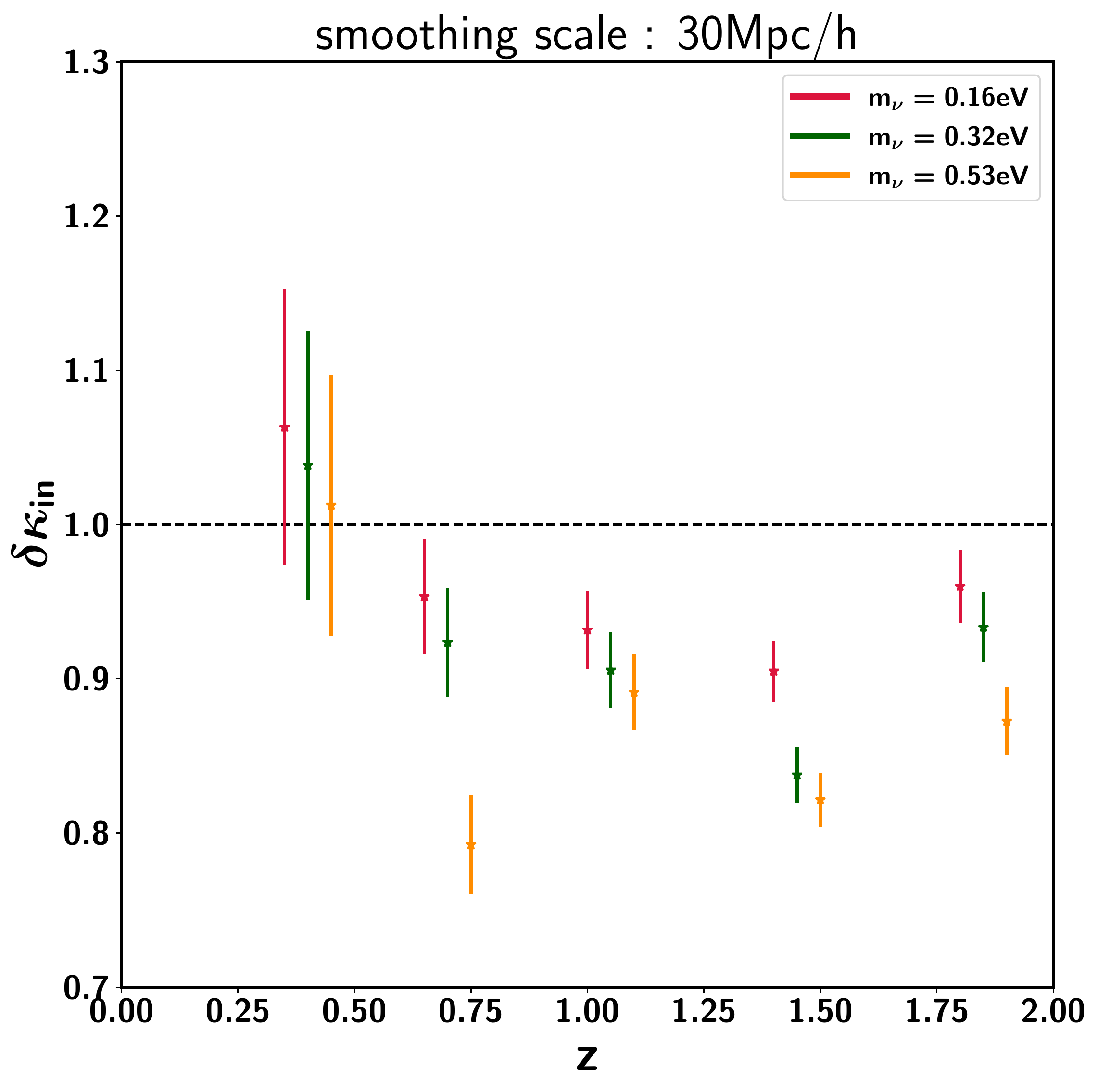}

\caption{{\it Top panel:} Redshift evolution of the lensing profile and imprint on CMB convergence map. All voids catalogues considered in this work are divided in different redshift bins for the different massive neutrinos cosmologies analysed here, $m_\nu=$0 eV (blue), $m_\nu=$0.16 eV (red), $m_\nu=$0.32 eV (green) and $m_\nu=$0.53 eV (yellow). {\it Bottom panel:} corresponding sensitivity parameter in  different redshift bin. From left to right, a different smoothing scale (10, 20, 30 $h^{-1}$Mpc) is considered for the void finder. }
\label{fig:zevol}
\end{figure*}

We then divide our void catalogues in different redshift bins from $z=0.2$ to $z=2$ and apply our stacking methodology to each bin, combining both CDM and neutrino maps. We show in the top panel of Figure~\ref{fig:zevol} the profiles measured for the different smoothing scales, while in the bottom panels we show $\delta\kappa_{in}$ as a function of redshift for the different massive neutrino cosmologies and the different smoothing scales. 
At low redshifts, although neutrinos will fall in large potential wells, we expect the smaller fluctuations to be smoothed. Namely, from Figure~\ref{fig:FS} we can see that as the redshift decreases, smaller scales will be affected. On the other hand, as we increase the redshift, the difference in the void population is also increasing when neutrinos are more massive (see Figure~\ref{fig:radius_dist}); the amount of small voids will be larger in the massless neutrino simulations with respect to the massive ones. These small structures would be more smoothed in their centre due to their size, on similar scales for which massive neutrinos will smooth the matter field. However, in the medium redshift range, the sensitivity parameter decreases with the neutrino mass. This is consistent with the slight tension claimed by \cite{hang2021}, where a lower signal in the observation appears in the DESI Legacy survey observation at $0.6 < z < 0.8$.

\subparagraph{Void radius evolution:}
Previously, we have measured that medium redshift ranges are showing more differences in void lensing imprints (Figure~\ref{fig:zevol}). In addition to that, it is also possible to prune the void catalogues in order to select voids that show a stronger lensing imprint. Moreover, as explained before and confirmed above, we expect the smaller voids to be more affected by the presence of massive neutrino since neutrino free-streaming will reduce the clustering at the scales corresponding to their sizes, making them less underdensed. Consequently, similarly to what we did in Sect.~\ref{sec:neutr_contr}, we split the void catalogues into different radius bins and measure the stacked lensing signal of all the sub-samples, separately.

\begin{figure*}

\includegraphics[width=.32\textwidth]{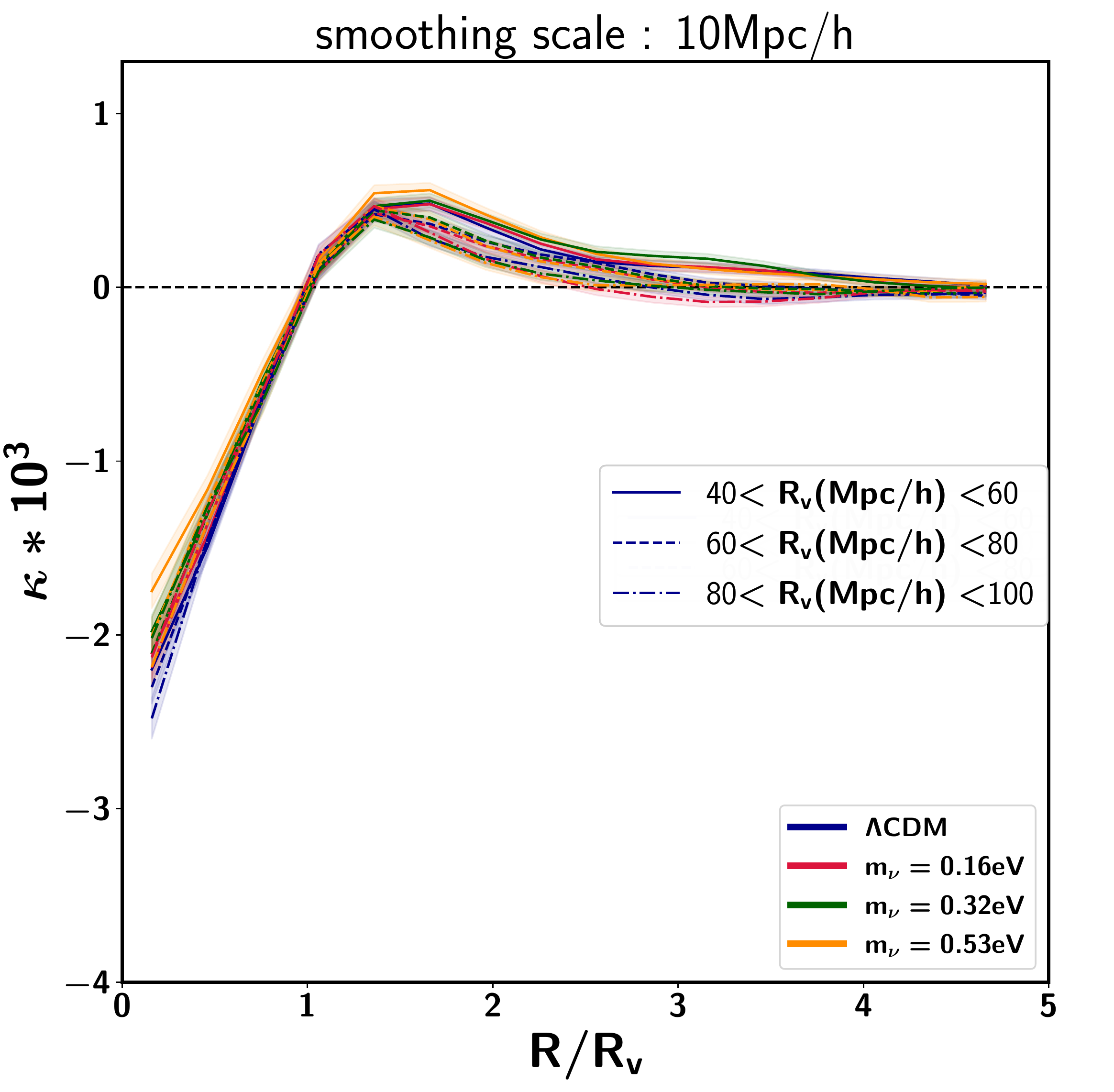}
\includegraphics[width=.32\textwidth]{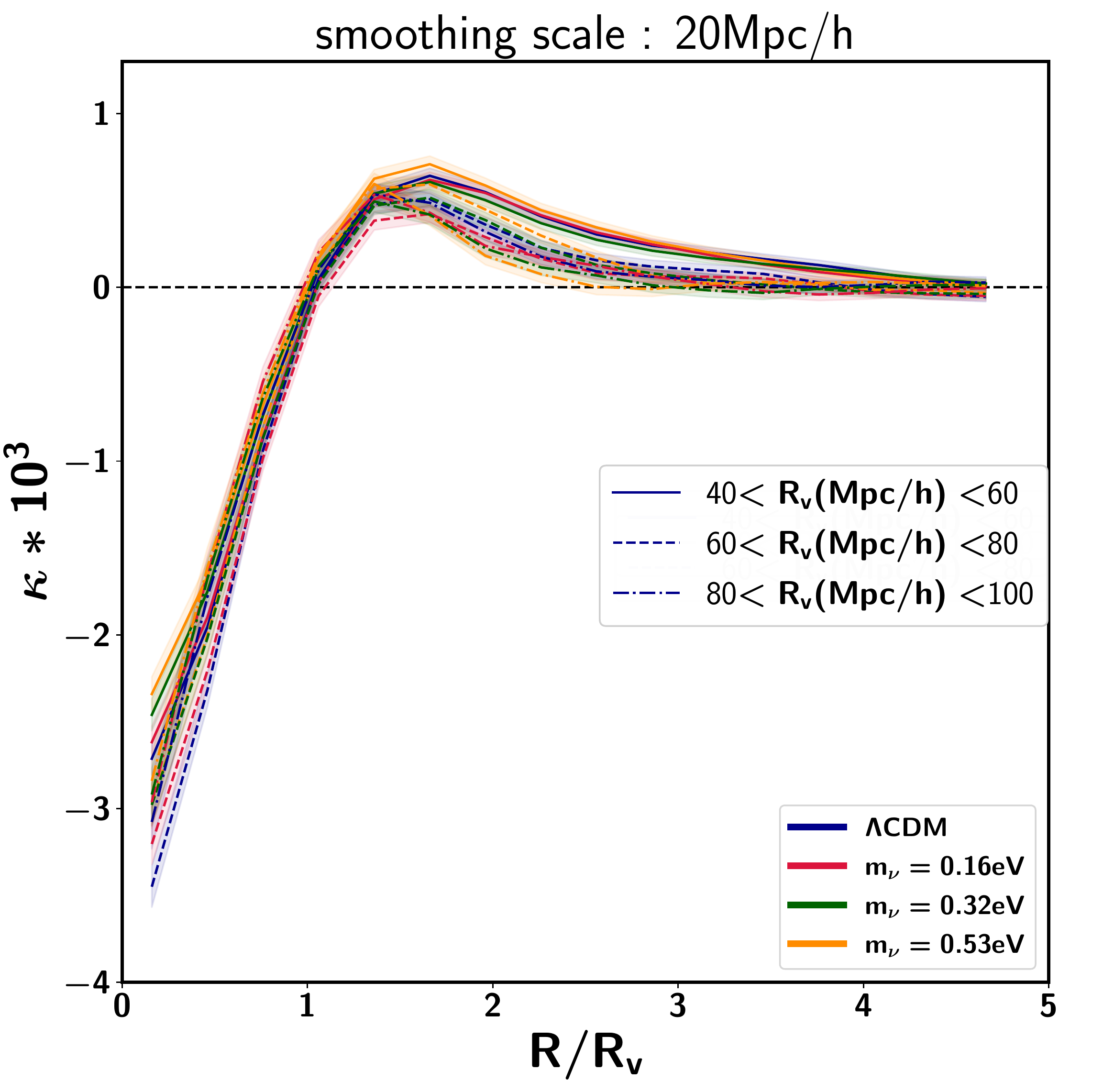}
\includegraphics[width=.32\textwidth]{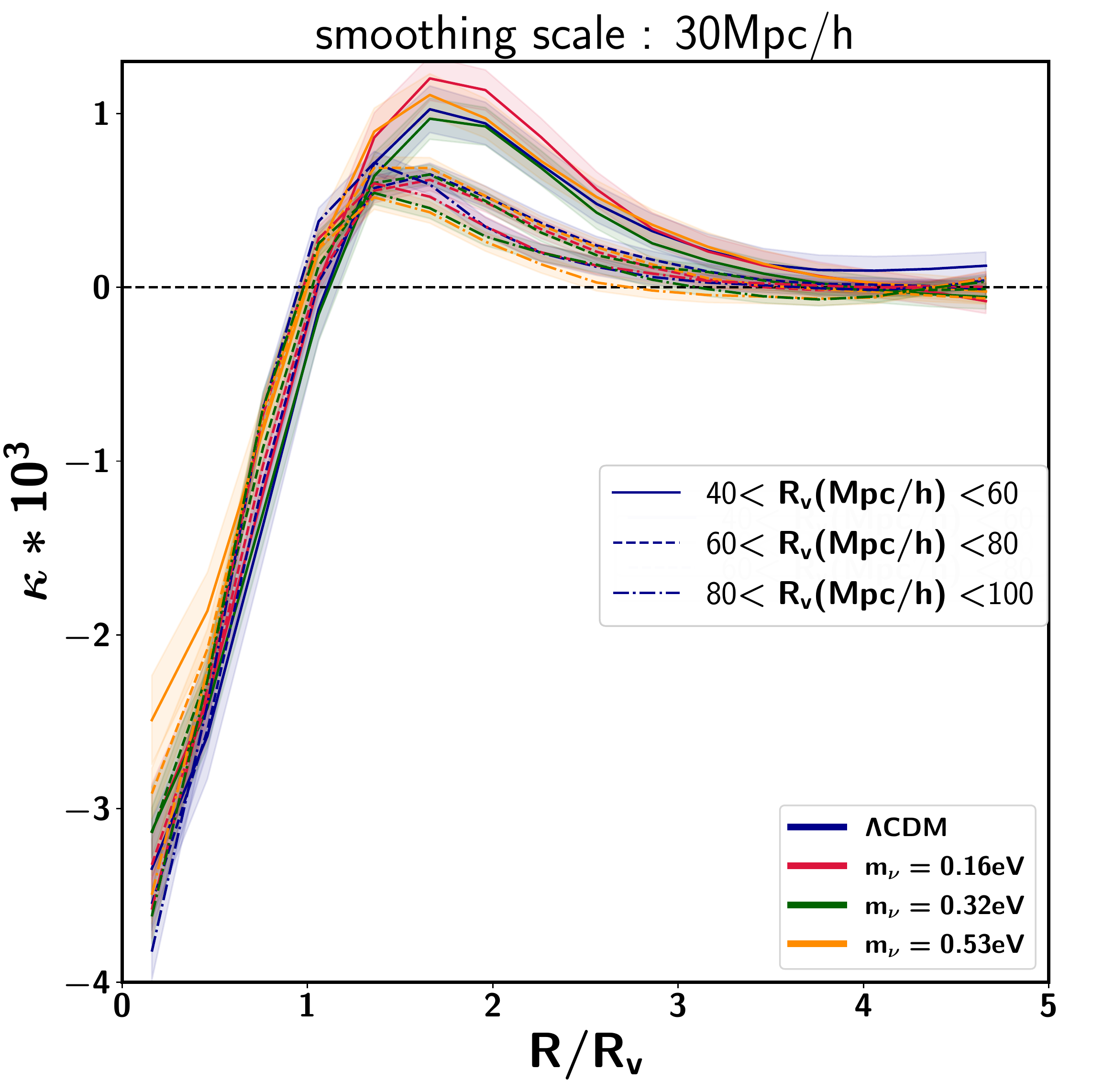}\quad

\includegraphics[width=.32\textwidth]{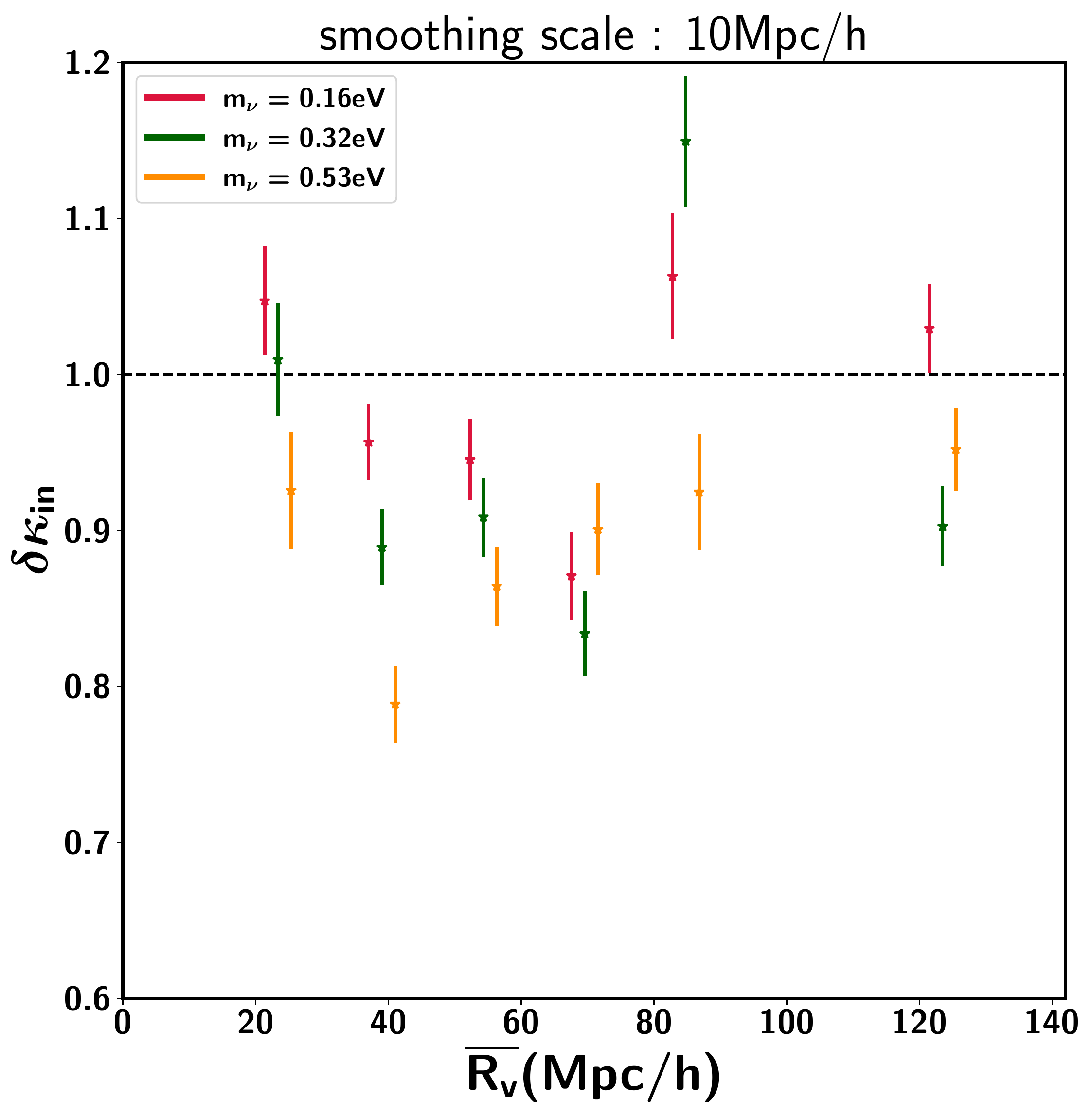}
\includegraphics[width=.32\textwidth]{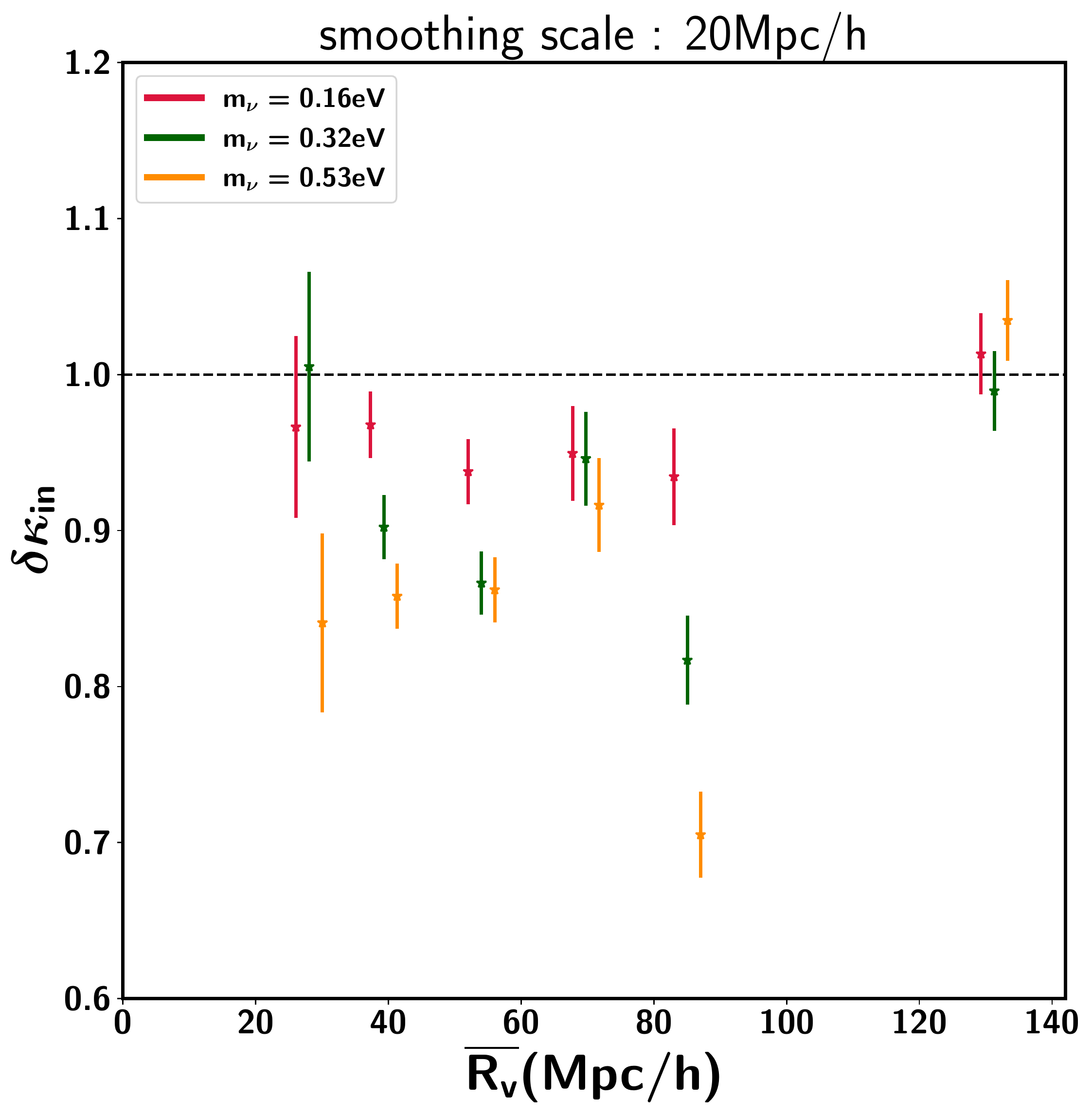}
\includegraphics[width=.32\textwidth]{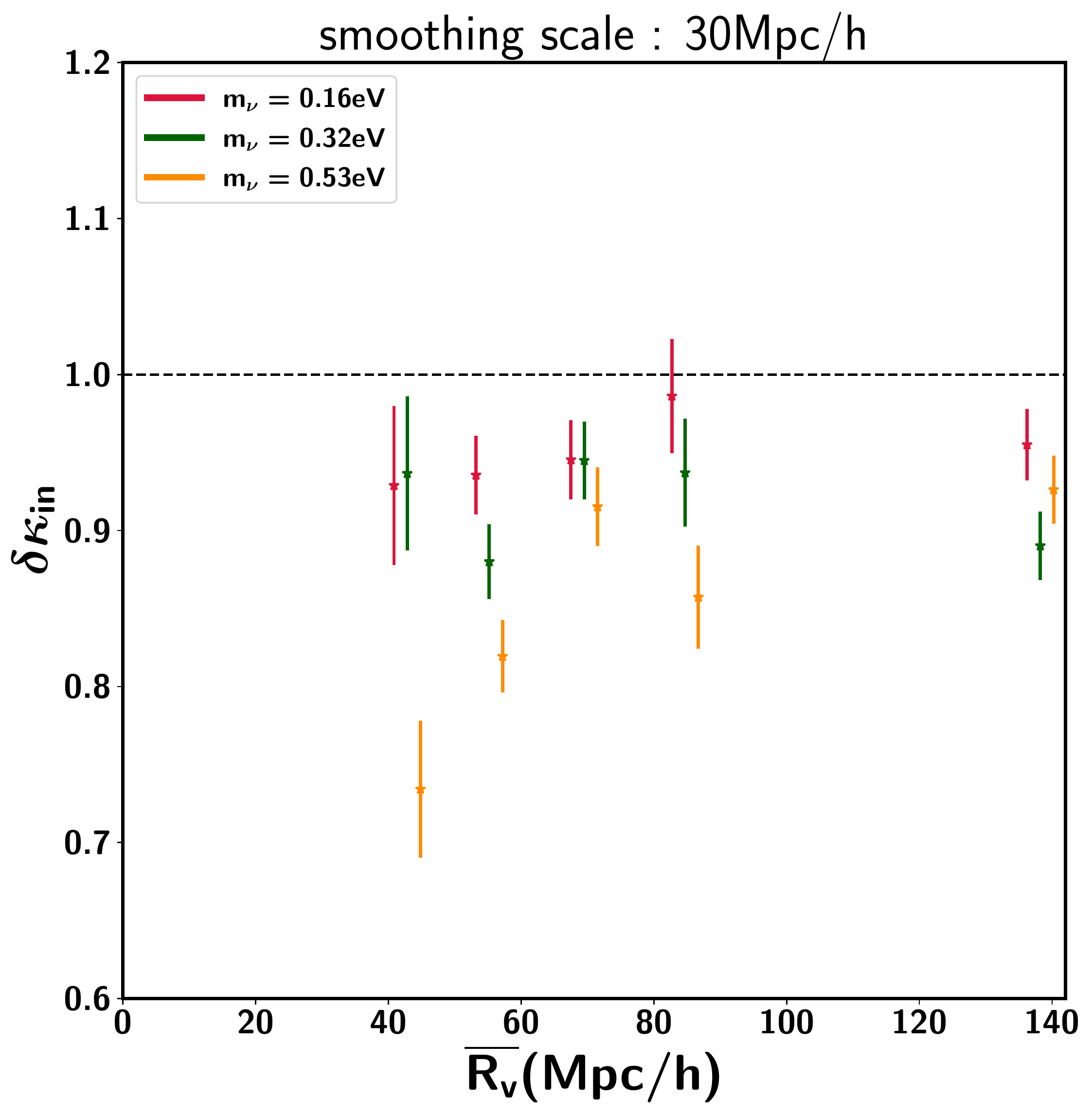}

\caption{Radius evolution of the void imprint on the CMB convergence map and lensing profile. All voids catalogues considered in this work are divided in different radius bins for the different massive neutrinos cosmologies; same description as Figure~\ref{fig:zevol}.}
\label{fig:Revol.pdf}
\end{figure*}

Figure~\ref{fig:Revol.pdf} shows the different void CMB lensing profiles (top panels) and the evolution of the sensitivity parameter as a function of the void radius (bottom panels), for the different neutrino masses and smoothing scales considered. Smaller voids have less pronounced lensing signal than medium radius voids, and the relative difference in the massive neutrino cosmologies seems to be enhanced as the object radius decreases and the smoothing scale increases. However, we note that - in particular for the lowest smoothing scale of 10 $h^{-1}$Mpc - the neutrino imprint seems to decrease; this maybe related to the fact that for small smoothing scales matter underdensities within the voids are less nonlinear, that is with smaller amplitude, and they become more linear as the neutrino mass increases.

Finally, we went one step further in Appendix \ref{sec:appendix_B} where we have reiterated the correlation measurement applying a double binning in void radius and redshift.

\section{Conclusions}
The aim of this work is to study the cosmic void imprint on the CMB-convergence in massive neutrinos cosmologies. To this purpose, we identify two ways in which massive neutrinos can alter this CMB void-lensing signal. First, massive neutrinos can induce selection effects in the void identification process as they affect the density of tracers used to identify them. Second, massive neutrinos suppress matter density perturbations inside cosmic voids, implying that void-lensing signals from voids with similar size might differ according to the neutrino mass considered.  In order to fully understand these two effects, we analysed how the presence of massive neutrinos modifies the void finding process, and thus the intrinsic properties of the void catalogues. Moreover, we studied the cross-correlation signal of cosmic voids with CMB lensing. For this work, we exploited a set of N-body cosmological simulations, the DEMNUni suite, in various massive neutrino scenarios (0 eV, 0.16 eV, 0.32 eV and 0.53 eV).

CMB-convergence full-sky maps were built in Born approximation via ray-tracing of CMB-photons moving across the dark matter distribution (CDM and neutrinos) of the DEMNuni simulations, combined with a stacking technique~\cite{Fabbian_2018,Hilbert2020} of comoving particle snapshots . With the same stacking technique, for each considered cosmology, halo catalogues in comoving snapshots, obtained from the DEMNUni simulations via a FoF algorithm, were organised in a full-sky lightcone. In each lightcone of DM haloes, we identified 2D voids using the void finder presented in~\cite{Carles_void}, and applied different levels of Gaussian smoothing to the density field, in order to probe void catalogues with different properties (such as the mean void radius or voids densities). By looking at the intrinsic void features, we have shown that: \begin{itemize}
    \item in terms of their abundances, shown in Figure~\ref{fig:radius_dist}, the presence of massive neutrinos tends to decrease the total number density of voids traced by haloes with a minimum mass of $M_h=2.5 \times 10^{12} h^{-1}M_\odot$. We note that the number density of small void radius tends to be more reduced than the large radius ones. Such a behaviour in the void size function can be directly explained by the decrease in the tracer density, due to structure formation suppression by free-streaming neutrinos, which in turn induces a merging of the smaller structures in the void identification procedure. As we increase the Gaussian smoothing in the void finder - and thus trace larger structures - we can reduce this effect.

    \item with regards to void profiles traced by the halo-void cross-correlation, we observe in Figure~\ref{fig:dens_prof_20} that voids in massive neutrino cosmologies seem to be slightly deeper than in the massless neutrino case. In fact, massive neutrinos slow down structure formation and therefore the overall density of haloes with mass larger than $M_h=2.5 \times 10^{12} h^{-1}M_\odot$ is decreased. Consequently, the density of halo tracers inside voids is reduced as well, making void profiles, traced by the halo-void cross-correlation, to look deeper in massive neutrino cosmologies.
\end{itemize} 
In this respect, 2D voids should be treated as potential tools to constrain neutrino masses, by looking at the void size function and the void-halo cross-correlation.

Besides that, in this work we consider also the void-CMB lensing cross-correlation and show that it can be considered as well an important observable to probe the neutrino mass. In fact, contrary to void-halo clustering, this observable has only a linear dependence on the void bias (which in turn depends on the population of haloes used to identify the voids), while it is independent from the halo bias, since CMB convergence maps depend directly on the underlying matter field (both CDM and neutrinos). Indeed, the void-CMB lensing cross-correlation could be used together with void-void clustering to constrain the void bias. In this work, we looked at two kinds of void-CMB lensing cross-correlations: the cross-correlation between cosmic voids and  CMB lensing deflections caused by the CDM field alone (i.e. void profiles traced by void-$\kappa_{\rm CDM}$ cross-correlation), and the cross-correlation between voids and CMB lensing deflections caused by the the neutrino field alone (i.e. void profiles traced by void-$\kappa_{m_\nu}$ cross-correlation), for the case of a high neutrino mass of $m_\nu=0.53$ eV. In fact, as voids are large structures (especially the one traced by haloes with quite a large mass) we expect neutrino free-streaming to be less effective than in galaxy clusters, which are characterised by smaller scales. We measured an anti-correlation signal in both neutrino and CDM convergence maps within voids.
In Figure~\ref{fig:neutrinocontrib_im}  we observe first a large increase in the amplitude of the anti-correlation signals (both for CDM and $\nu$) for increasing smoothing scales; this can be explained by the fact that with our void-finder we are identifying density fluctuations that in amplitude are larger than the smoothing threshold chosen in the void-finder. If this happens at large scales (i.e. large smoothing scales), this means that the identified density fluctuation is already nonlinear at those scales, and therefore we will observe even more nonlinear fluctuations on scales smaller than the smoothing one, scales which enclosed within the void profile. Therefore, in terms of nonlinear underdensities, we will observe deeper voids as the smoothing scale increases, as shown in Figure~\ref{fig:neutrinocontrib_im} for the unbinned total signal and $m_\nu=0.53$ eV, and in Figure~\ref{fig:Revol.pdf} for different radius bins and neutrino masses.

In addition, in Figure~\ref{fig:neutrinocontrib_im} we can also observe that the ratio, $\kappa_{m_\nu} /\kappa_{CDM}$, i.e. between the void profiles in the neutrino and CDM fields respectively, increases with the smoothing scale. However, the signal is noisy and larger resolution simulations will be needed to study this effect in detail. We leave it for future work.

Since the effects of massive neutrinos depend on the scales and redshifts considered, we split our catalogues into various subsets of voids of similar sizes at different redshifts, in order to measure the void-lensing profiles while these parameters evolve. In Figure~\ref{fig:Zevol_separate.pdf} for the smallest smoothing scale we observe a slight increase of $\kappa_{m_\nu}$/$\kappa_{\rm CDM}$ as the redshift increases. The trend is inverted for the largest smoothing scale. However, this behaviour is hardly distinguishable due to the large errorbars and again we leave it for future work. Analogously, comparing the left and right panels of Figure~\ref{fig:Zevol_separate.pdf}, we observe that increasing the smoothing scale inverts the redshift evolution of the void-lensing profiles: larger voids are deeper for larger redshifts (high-$z$ voids will enclose more nonlinear perturbations for larger smoothing scales, boosting thus the anti-correlation amplitude) and smaller smoothing scale voids will be deeper for smaller redshifts (the smoothing technique selection affects only very small scales and we can observe the non linear growth of underdensities via gravitational instability).

Moreover, we have measured different correlation signals between voids and CMB convergence maps for the full void sample as well as for various sub-samples, at different redshifts in different neutrino cosmologies. 
For the considered void populations (traced by haloes with mass larger than $M_h=2.5\times 10^{12} h^{-1}M_\odot$), we note in Figure~\ref{fig:FULL_SIGNAL} that the presence of massive neutrinos tends to decrease the amplitude of the void-CMB lensing cross-correlation w.r.t. the massless neutrino case, i.e. to produce  shallower void profiles. We note that this effect is more enhanced as the neutrino mass increases. This can be explained by the theory of cosmological perturbations in the presence of massive neutrinos which suppress the nonlinear evolution of matter density perturbations, both overdensities and underdensities (such as voids). Therefore, in massive neutrinos scenarios the lensing convergence amplitude is suppressed below the free-streaming scale and consequently, for such scales, void-lensing profiles will be less deep as the neutrino mass increases. Similar trends are also showed in Figures~\ref{fig:zevol}-\ref{fig:Revol.pdf} as functions of redshift and bin in radius. Worth of note, these results could point neutrinos as the explanation for the claim from recent observation campaigns~\cite{hang2021,kovacs2022} of shallower void profiles measured via the cross-correlation between voids and CMB lensing. 

Our simulated measurements show the potential power of using in future particular setups in the void identification pipeline (eg larger structures) to increase the void sensitivity as a neutrino mass probe. Indeed, one of the main results of this work is represented by the trend of the sensitivity parameter, $\delta \kappa_{in}$ shown in Figure~\ref{fig:FULL_sensiv}. It shows that exploiting a large smoothing scale in the void search could help in distinguish very clearly between different massive neutrino cosmological scenarios.

Finally, we can claim that we observed a clear dependence on the neutrino mass in the void-lensing signal when the full void catalogue is considered, and this is in particular encouraging for the next-generation surveys, that will provide a unprecedentedly large catalogues of cosmic voids up to redshift $z=2$.
A possible extension  of this work could be to exploit simulated galaxy catalogues and to include observational systematics, in order to verify the ability of future surveys to detect the cosmological dependence on the neutrino mass in void-lensing correlation signals.




\begin{appendices}
\appendix
\appendix

\section{The massive neutrinos behaviour at different scales and redshifts}\label{sec:appendix_A}
\subparagraph{Redshift evolution:}\label{sec:redshift_evol_neutrinoeffect}

\begin{figure*}[h!]

\includegraphics[width=150mm]{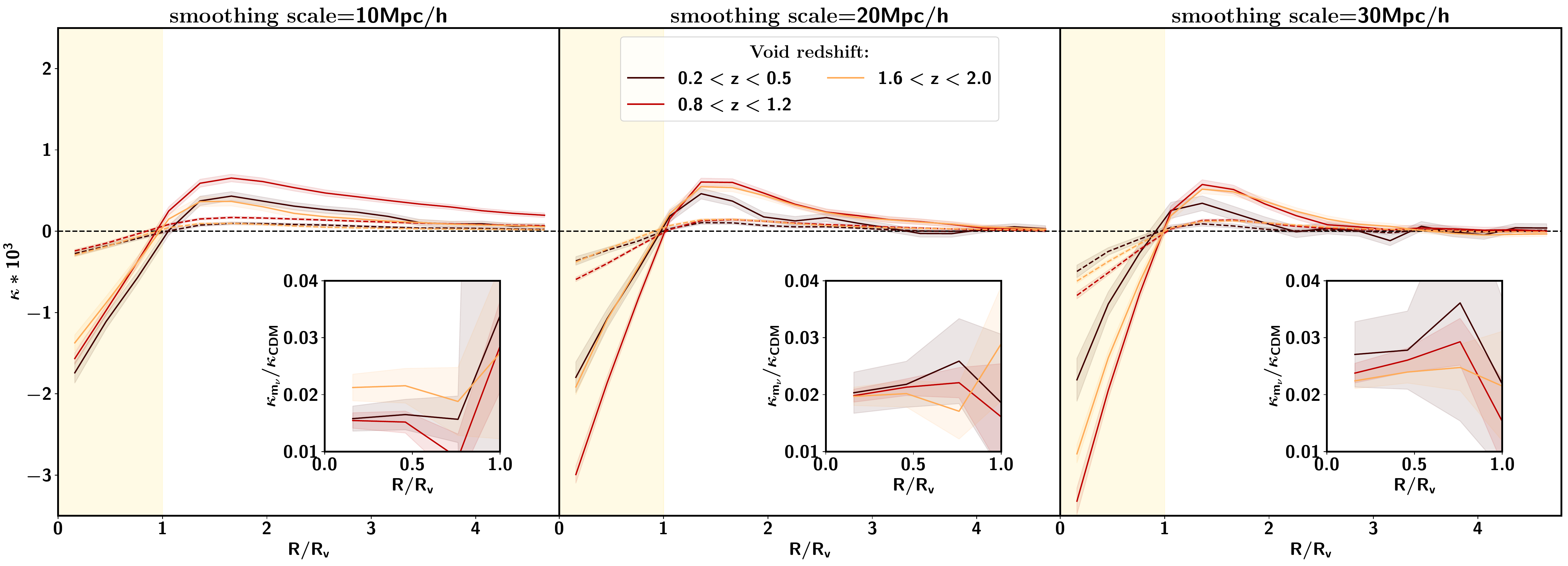}

\caption{Redshift evolution. Imprint caused by neutrinos {with $m_\nu=0.53$ eV} (dashed lines) and CDM only (solid line) for voids in three different redshift bins for the three smoothing scale 10 $h^{-1}$Mpc ({\it left panel}), 20 $h^{-1}$Mpc ({\it middle panel}), 30 $h^{-1}$Mpc ({\it right panel}). The shaded region represents the fluctuations of the signal measured in 1,000 randomly-generated CMB lensing map (see Section~\ref{sec:stacking_meth}). The insight plot in each panel is the ratio of the signal induced by neutrinos w.r.t. the one induced by CDM.}
\label{fig:Zevol_separate.pdf}
\end{figure*}

We want to verify whether or not massive neutrino present in cosmic voids are showing variation in their CMB imprints at different redshift, that is to say if the contribution of massive neutrino in the lensing imprints of voids evolves with time. To do so, we divide our sample in three different redshift bins ($0.2<z<0.5$; $0.8<z<1.2$ and $1.6<z<2.0$) and perform the stacking measurement of the signal in both neutrino-only lensing maps and CDM-only ones, for all the redshift sub-samples. Figure~\ref{fig:Zevol_separate.pdf} shows the separate imprints of neutrinos (dashed lines) and CDM (solid lines) in CMB lensing maps at different redshifts, for the three different smoothing scales. In the CDM only map, the amplitude of the correlation at the void centre seems to follow the behaviour of the CMB kernel with a peak at redshift $z\sim 1.5$. 
If we compare the difference in the signal within the three smoothing scales, in agreement with Figure~\ref{fig:neutrinocontrib_im}, we can observe that the redshift evolution of the lensing signal is more pronounced for both the CDM-only signal and the neutrino-only one as we increase the smoothing scale. In fact, if we look at Figure~\ref{fig:FS} and \ref{fig:central_dens_radii}, we see that the mean void radius (and most of the voids) in the 10 $h^{-1}$Mpc smoothing catalogue is below the free-streaming length considered here. Consequently, we do not expect a strong redshift evolution in the neutrinos lensing signal. However, once we increase the smoothing scale, the objects identified will reach sizes greater than the free-streaming length of neutrinos, thus these structures could be affected at different redshifts.
The insight plots of Figure~\ref{fig:Zevol_separate.pdf} show the ratio between CDM and neutrino contributions in the lensing signal at the void centre; however, the contribution signal is quite noisy, and we cannot observe a significant evolution in redshift for any of the smoothing scales used, with an average value for the ratio close to $2\%$. Given a specific neutrino mass, the lensing imprint directly caused by massive neutrino follows a similar redshift evolution as the imprint generated by CDM.

\begin{figure}[h!]

\includegraphics[width=150mm]{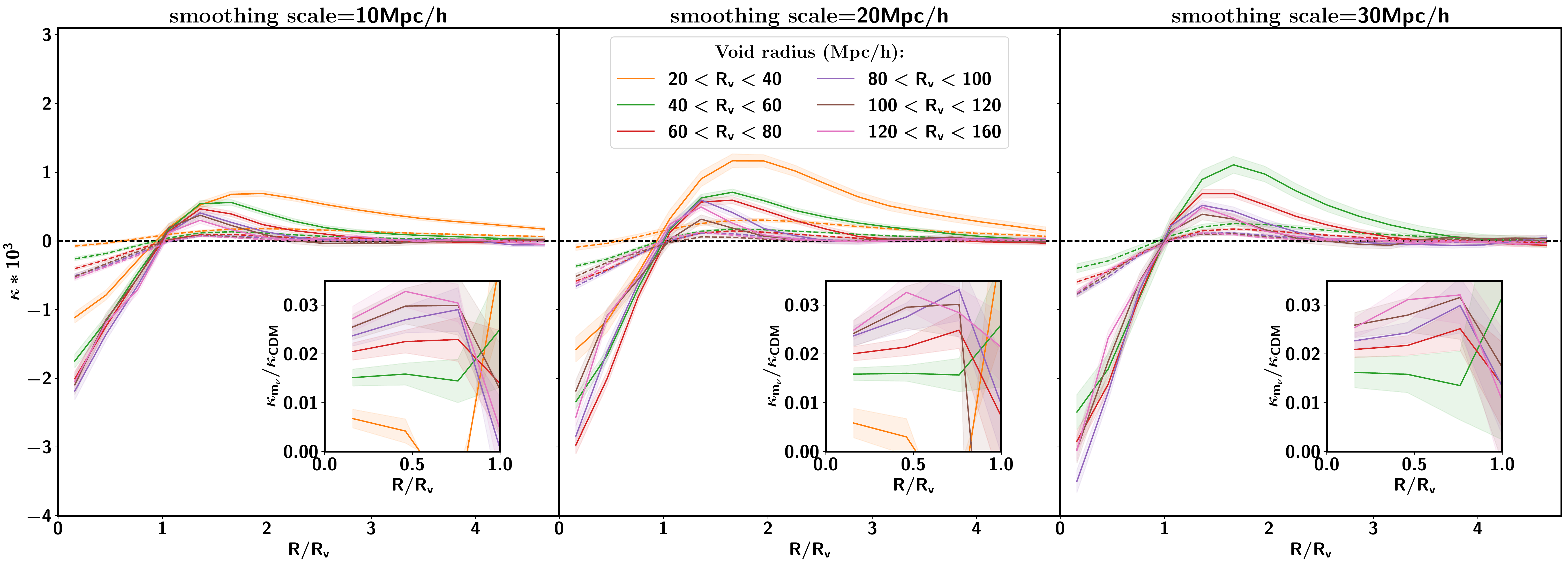}
\caption{Radius evolution. Imprint caused by neutrinos with $m_\nu=0.53$ eV (dashed lines) and CDM only (solid line) for voids in six different radius bins for the three different smoothing scales: 10 $h^{-1}$Mpc ({\it left panel}), 20 $h^{-1}$Mpc ({\it middle panel}), 30 $h^{-1}$Mpc ({\it right panel}). The shaded regions represent the fluctuations of the signal measured in 1,000 randomly-generated CMB lensing map (see Section~\ref{sec:stacking_meth}). The insight plot in each panel is the ratio of the signal induced by neutrinos w.r.t. the one induced by CDM.}
\label{fig:Revol_separate}
\end{figure}

\subparagraph{Radius evolution:}\label{sec:Radius_evol_neutrinoeffect}
The effects of massive neutrinos on the structures of the universe is two-fold. At scales smaller than the free-streaming scale $\lambda_{\rm FS}$, due to their velocities, the neutrinos will travel over the different fluctuations in the potential field without being affected,  smoothing in fact the inhomogeneities. At scales larger than $\lambda_{\rm FS}$, the neutrinos will fall into the gravitional potentials. Consequently, we do expect voids catalogues including scales similar or larger than $\lambda_{\rm FS}$ to be more {\it devoid} of neutrinos than the ones with smaller structures.
Thus, we want to investigate the abundance of massive neutrinos in underdensed structure as a function of their sizes: for this purpose, we have divided our voids catalogues in 6 radius bins, 
and we have measured the lensing signal in each of these sub-sampled catalogues by applying the stacking technique to both the lensing signal due to CDM particles and the lensing signal due to massive neutrinos\footnote{Note that we have discard the lower radius bin ($20<R_{\rm v}(h^{-1}{\rm Mpc})<40$) in the 30$h^{-1}$Mpc smoothing scale due the low numerosity of the sample.}. Results are shown in Figure~\ref{fig:Revol_separate} for the three different smoothing parameters (from left to right panels). A stronger "de-lensing" signal due to massive neutrinos is observed for larger voids: the neutrinos are, as expected, less present in the largest objects. The insight plots in the Figure show the ratio between the lensing signal due to CDM-only to massive neutrino-only at the void centre.  
The contribution of massive neutrinos on the lensing signal seems for all the cases to be stronger as one increases the radius of the lensed voids, for both massive neutrinos and CDM imprints. Such behaviours suggest that, as we decrease the void radius, cosmic voids identified in the matter field will be less underdensed with neutrinos, which is consistent with the fact that neutrino will fall in potential wells at scales larger than $\lambda_{\rm FS}$, while travelling through density fluctuations at smaller scales. 

\section{Combined binning, in redshift and radius}
\label{sec:appendix_B}
\begin{figure*}
\begin{center}

\includegraphics[width=1.\columnwidth]{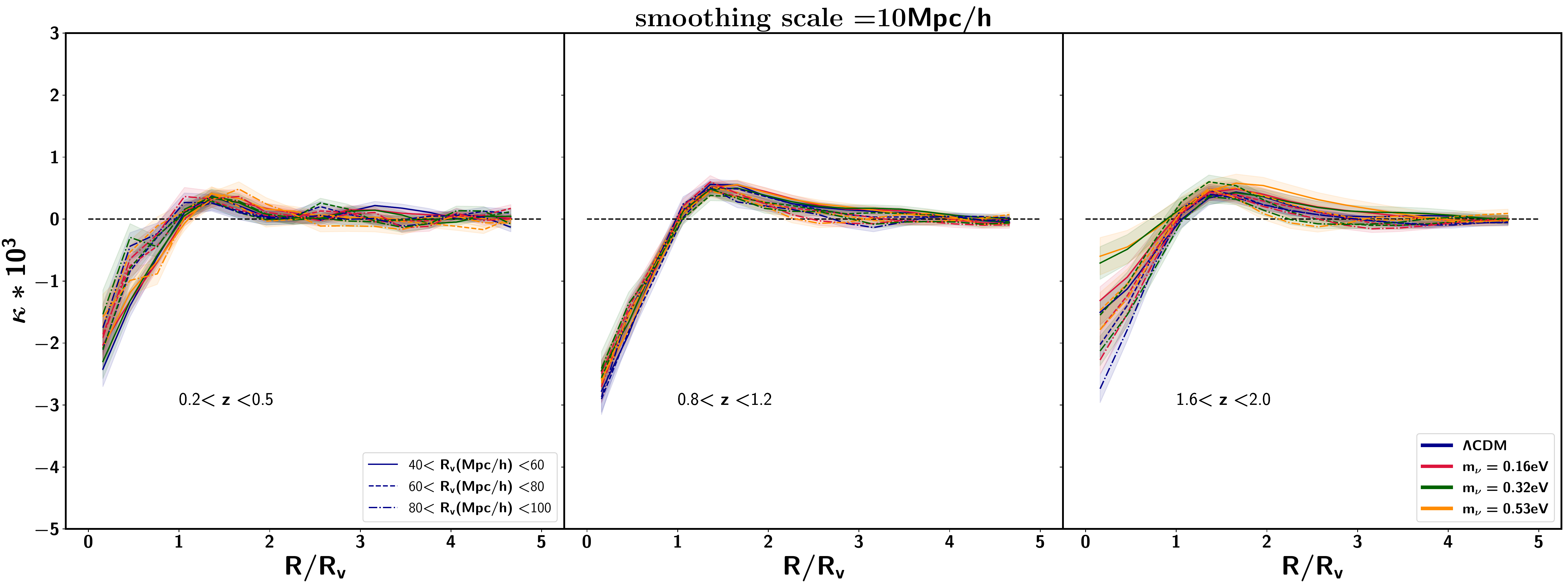}
\includegraphics[width=1.\columnwidth]{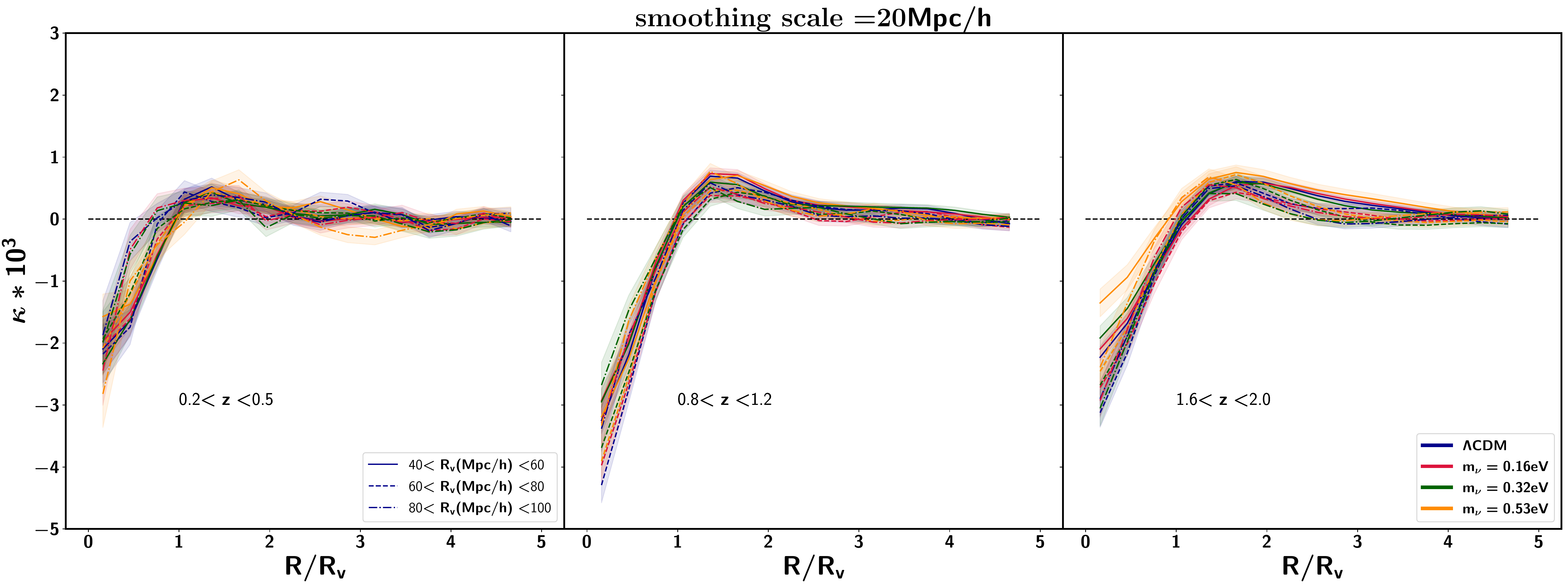}
\includegraphics[width=1.\columnwidth]{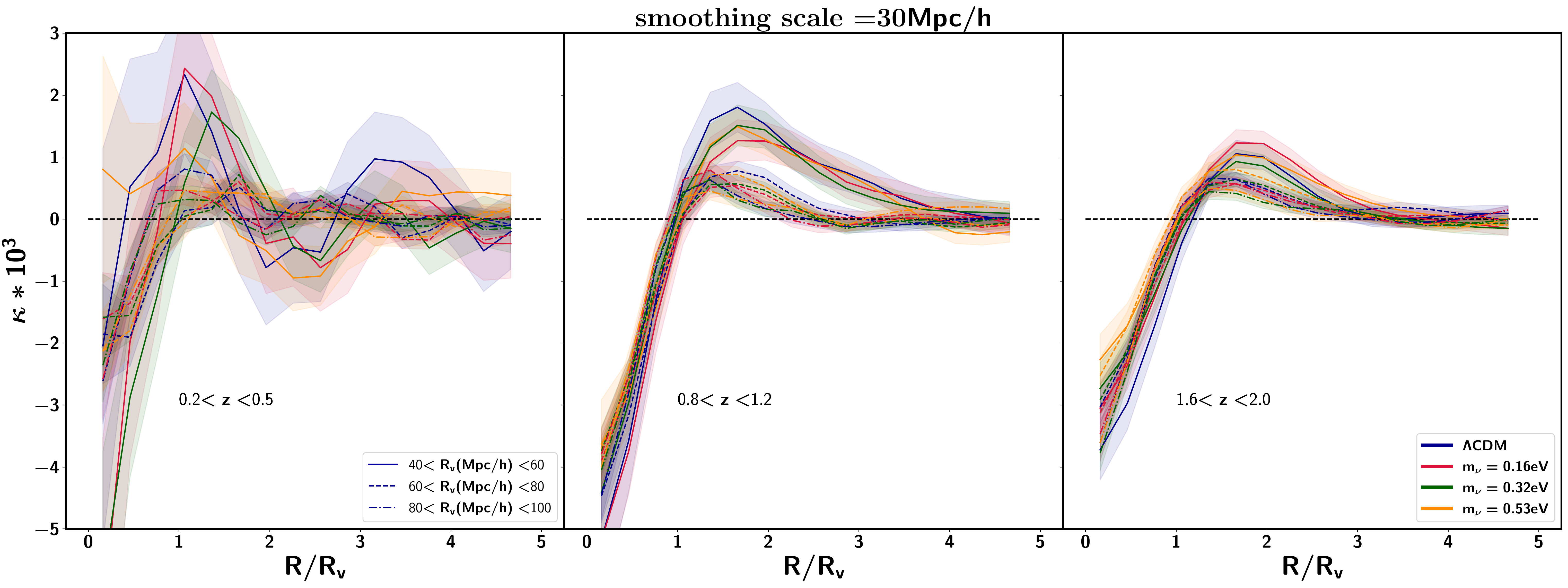}
\end{center}
\caption{Lensing imprint of cosmic voids, with a combined  binning in both redshift and radius, for the different smoothing scale (from top to bottom: 10, 20, 30 $h^{-1}$Mpc). The shade region is the error computed using the methodology described in Section~\ref{sec:stacking_meth}. Different line-styles refer to the radius bin, while the redshift evolution is shown from left to right panels. Colours discriminate between massive and massless neutrino cosmologies.}
\label{fig:finalevol_zoom.pdf}
\end{figure*}

\begin{figure*}
\begin{center}

\includegraphics[width=.32\textwidth]{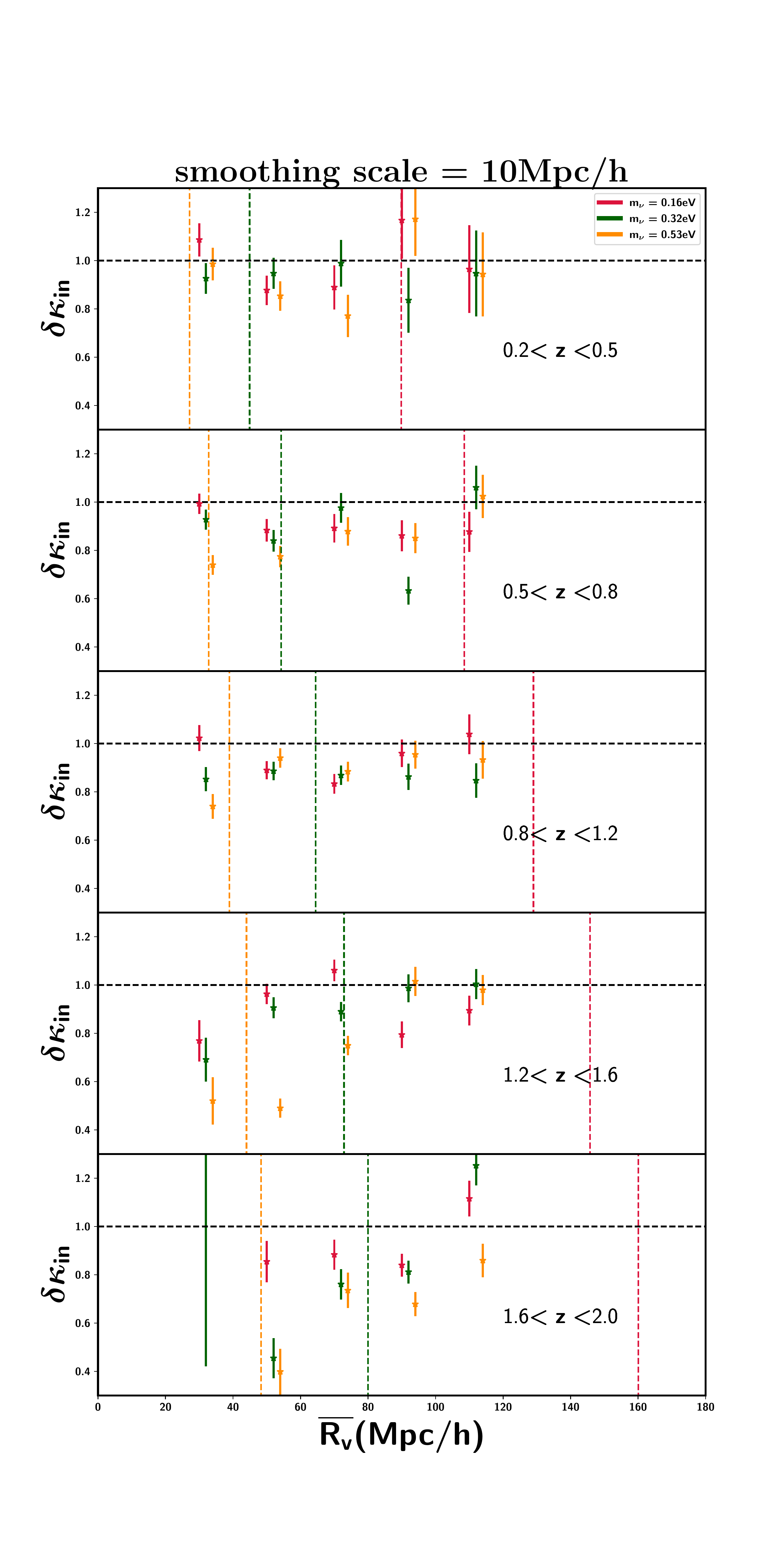}
\includegraphics[width=.32\textwidth]{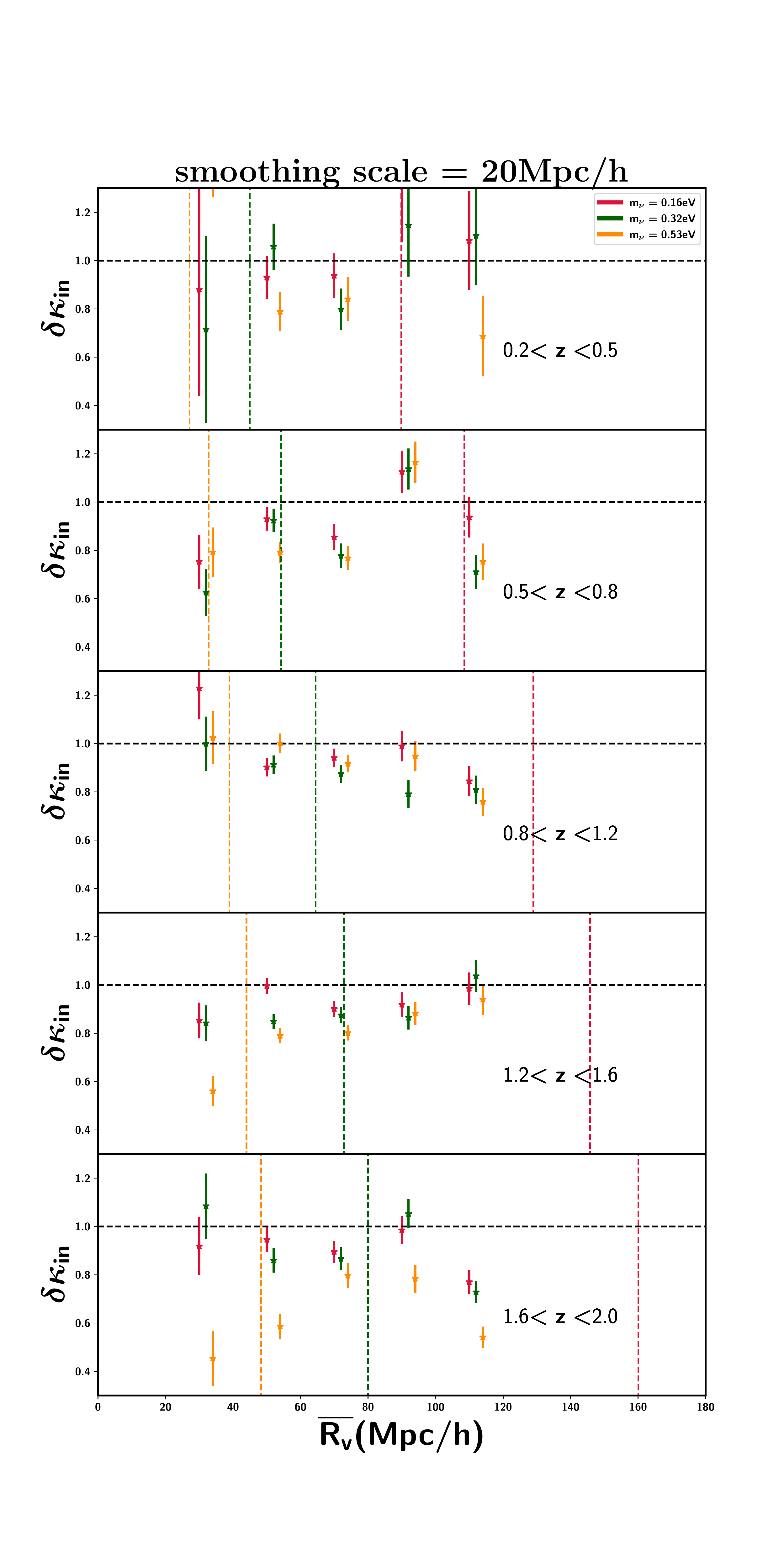}
\includegraphics[width=.32\textwidth]{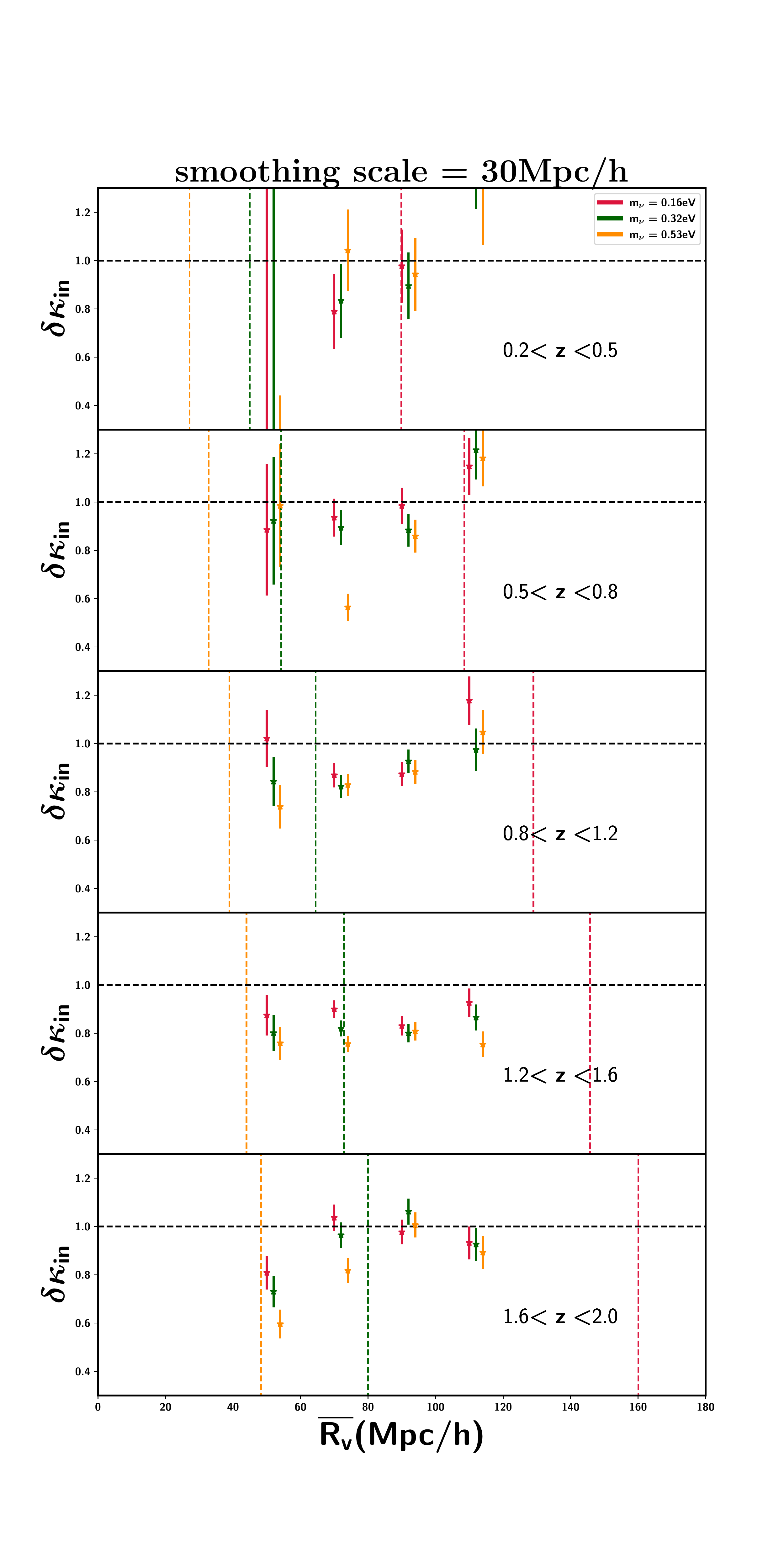}
\end{center}
\caption{Sensitivity parameter $\delta\kappa_{in}$ as a function of mean void radius for each redshift bin considered (from top to bottom); in each column the different smoothing scale considered. The vertical dashed lines represent the free-streaming length $\lambda_{\rm FS}$ of the neutrinos - Eq.~\eqref{eq:FS} - for the redshift bin.}
\label{fig:finalratio}
\end{figure*}

The lensing imprint detected in massive neutrino simulations w.r.t. the massless neutrinos $\Lambda$CDM model depends on the redshift and the radius of the lenses used. In this section, we intend to measure such signal in different sub-samples binned both in radius and redshift. We have  binned our sample in the same ranges in redshift and radius as previously, and proceed to the stacking of these sub-samples in the CMB lensing maps of our simulation. The lensing profiles and sensitivity evolution for the different bins and neutrino cosmologies are shown in Figure~\ref{fig:finalevol_zoom.pdf} and \ref{fig:finalratio}, respectively. We added intermediate bins in Figure~\ref{fig:finalratio} in order to look in details at the redshift and radius evolution of the sensitivity parameter. The vertical lines in Figure~\ref{fig:finalratio} correspond to the value of the free-streaming scale $\lambda_{\rm FS}$ (computed for the average redshift in the bin). We note that in the lower redshift range, the number of stacked objects decreases significantly, and thus the correlation profiles becomes too noisy to disentangle the mass of neutrino species. However, in the higher redshift bins, due to the increase in the observed volume, we reach a number of voids higher enough to appreciate differences in the stacked profiles among the different cosmologies. In particular, we note that for the higher redshift bins, small and medium size voids show a relatively lower CMB lensing imprint once one considers massive neutrinos. On the contrary, larger structures (scales of the order of $\sim$ 100 $h^{-1}$Mpc) are showing sensitivity parameters close to $1$, suggesting no significant imprints in the correlation profile due to neutrinos.  

\end{appendices}

\section*{Acknowledgements}
We warmly thank Giovanni Verza for very useful comments. MC is partially supported by a 2021-2022 Research and Education grant from Fondazione CRT. GF acknowledges the support of the European Research Council under the Marie Sk\l{}odowska Curie actions through the Individual Global Fellowship No.~892401 PiCOGAMBAS.  The OAVdA is managed by the Fondazione Cl\'ement Fillietroz-ONLUS, which is supported by the Regional Government of the Aosta Valley, the Town Municipality of Nus and the Unit\'e des Communes valdotaines Mont-\'Emilius.
Part of this research used resources of the National Energy Research Scientific Computing Center (NERSC), a U.S. Department of Energy Office of Science User Facility operated under Contract No. DE-AC02-05CH11231.
The DEMNUni simulations were carried out in the framework of ``The Dark Energy and Massive-Neutrino Universe" project, using the Tier-0 IBM BG/Q Fermi machine and the Tier-0 Intel OmniPath Cluster Marconi-A1 of the
Centro Interuniversitario del Nord-Est per il Calcolo Elettronico (CINECA). We acknowledge
a generous CPU and storage allocation by the Italian Super-Computing Resource Allocation
(ISCRA) as well as from the HPC MoU CINECA-INAF, together with storage from INFN-CNAF and INAF-IA2.


\bibliographystyle{JHEP_old}		  

\bibliography{biblio}

\label{lastpage}

\end{document}